\documentclass[twocolumn,prd,superscriptaddress,altaffilletter,nofootinbib]{revtex4-1}
\usepackage{amsmath}
\usepackage{amssymb}
\usepackage{amsfonts}
\usepackage{comment}
\usepackage{graphicx}
\usepackage{multirow}
\usepackage{hyperref}
\usepackage[nolist]{acronym}
\usepackage[usenames,dvipsnames]{xcolor}
\usepackage{xspace}
\usepackage{tikz}
\usepackage[squaren]{SIunits}
\usepackage{textcomp}
\usepackage[caption=false]{subfig}
\usepackage[export]{adjustbox}
\usetikzlibrary{arrows,shapes,trees,decorations.pathreplacing}
\allowdisplaybreaks

\newcommand{\mchirp}{\ensuremath{\mathcal{M}}\xspace}
\newcommand{\msun}{\ensuremath{\mathrm{M}_{\odot}}\xspace}
\newcommand{\Mc}{\ensuremath{\mathcal{M}}}

\begin{document}

\title{Template bank for spinning compact binary mergers in the second observation run of Advanced LIGO and the first observation run of Advanced Virgo}

\author{Debnandini Mukherjee}
\email{debnandini.mukherjee@ligo.org}
\affiliation{Leonard E.\ Parker Center for Gravitation, Cosmology, and Astrophysics, University of Wisconsin-Milwaukee, Milwaukee, WI 53201, USA}
\affiliation{Indian Institute of Technology Bombay, India}
\affiliation{Department of Physics, The Pennsylvania State University, University Park, PA 16802, USA}

\author{Sarah Caudill}
\email{sarah.caudill@ligo.org}
\affiliation{Leonard E.\ Parker Center for Gravitation, Cosmology, and Astrophysics, University of Wisconsin-Milwaukee, Milwaukee, WI 53201, USA}
\affiliation{Nikhef, Science Park, 1098 XG Amsterdam, Netherlands}

\author{Ryan Magee}
\affiliation{Department of Physics, The Pennsylvania State University, University Park, PA 16802, USA}
\affiliation{Institute for Gravitation and the Cosmos, The Pennsylvania State University, University Park, PA 16802, USA}

\author{Cody Messick}
\affiliation{Department of Physics, The Pennsylvania State University, University Park, PA 16802, USA}
\affiliation{Institute for Gravitation and the Cosmos, The Pennsylvania State University, University Park, PA 16802, USA}

\author{Stephen Privitera}
\affiliation{Albert-Einstein-Institut, Max-Planck-Institut f{\"u}r Gravitationsphysik, D-14476 Potsdam-Golm, Germany}

\author{Surabhi Sachdev}
\affiliation{LIGO Laboratory, California Institute of Technology, MS 100-36, Pasadena, California 91125, USA}
\affiliation{Department of Physics, The Pennsylvania State University, University Park, PA 16802, USA}

\author{Kent Blackburn}
\affiliation{LIGO Laboratory, California Institute of Technology, MS 100-36, Pasadena, California 91125, USA}

\author{Patrick Brady}
\affiliation{Leonard E.\ Parker Center for Gravitation, Cosmology, and Astrophysics, University of Wisconsin-Milwaukee, Milwaukee, WI 53201, USA}

\author{Patrick Brockill}
\affiliation{Leonard E.\ Parker Center for Gravitation, Cosmology, and Astrophysics, University of Wisconsin-Milwaukee, Milwaukee, WI 53201, USA}

\author{Kipp Cannon}
\affiliation{Canadian Institute for Theoretical Astrophysics, 60 St. George Street, University of Toronto, Toronto, Ontario, M5S 3H8, Canada}
\affiliation{RESCEU, The University of Tokyo, Tokyo, 113-0033, Japan}

\author{Sydney J. Chamberlin}
\affiliation{Department of Physics, The Pennsylvania State University, University Park, PA 16802, USA}
\affiliation{Institute for Gravitation and the Cosmos, The Pennsylvania State University, University Park, PA 16802, USA}

\author{Deep Chatterjee}
\affiliation{Leonard E.\ Parker Center for Gravitation, Cosmology, and Astrophysics, University of Wisconsin-Milwaukee, Milwaukee, WI 53201, USA}

\author{Jolien D. E. Creighton}
\affiliation{Leonard E.\ Parker Center for Gravitation, Cosmology, and Astrophysics, University of Wisconsin-Milwaukee, Milwaukee, WI 53201, USA}

\author{Heather Fong}
\affiliation{Canadian Institute for Theoretical Astrophysics, 60 St. George Street, University of Toronto, Toronto, Ontario, M5S 3H8, Canada}

\author{Patrick Godwin}
\affiliation{Department of Physics, The Pennsylvania State University, University Park, PA 16802, USA}
\affiliation{Institute for Gravitation and the Cosmos, The Pennsylvania State University, University Park, PA 16802, USA}

\author{Chad Hanna}
\affiliation{Department of Physics, The Pennsylvania State University, University Park, PA 16802, USA}
\affiliation{Department of Astronomy and Astrophysics, The Pennsylvania State University, University Park, PA 16802, USA}
\affiliation{Institute for Gravitation and the Cosmos, The Pennsylvania State University, University Park, PA 16802, USA}

\author{Shasvath Kapadia}
\affiliation{Leonard E.\ Parker Center for Gravitation, Cosmology, and Astrophysics, University of Wisconsin-Milwaukee, Milwaukee, WI 53201, USA}

\author{Ryan N.\ Lang}
\affiliation{Hillsdale College, Hillsdale, MI 49242, USA}

\author{Tjonnie G. F. Li}
\affiliation{Department of Physics, The Chinese University of Hong Kong, Shatin, New Territories, Hong Kong}

\author{Rico K. L. Lo}
\affiliation{Department of Physics, The Chinese University of Hong Kong, Shatin, New Territories, Hong Kong}
\affiliation{LIGO Laboratory, California Institute of Technology, MS 100-36, Pasadena, California 91125, USA}

\author{Duncan Meacher}
\affiliation{The Pennsylvania State University, University Park, PA 16802, USA}

\author{Alex Pace}
\affiliation{The Pennsylvania State University, University Park, PA 16802, USA}

\author{Laleh Sadeghian}
\affiliation{Leonard E.\ Parker Center for Gravitation, Cosmology, and Astrophysics, University of Wisconsin-Milwaukee, Milwaukee, WI 53201, USA}

\author{Leo Tsukada}
\affiliation{RESCEU, The University of Tokyo, Tokyo, 113-0033, Japan}
\affiliation{Department of Physics, Graduate School of Science, The University of Tokyo, Tokyo, 113-0033, Japan}

\author{Leslie Wade}
\affiliation{Department of Physics, Hayes Hall, Kenyon College, Gambier, Ohio 43022, USA}

\author{Madeline Wade}
\affiliation{Department of Physics, Hayes Hall, Kenyon College, Gambier, Ohio 43022, USA}

\author{Alan Weinstein}
\affiliation{LIGO Laboratory, California Institute of Technology, MS 100-36, Pasadena, California 91125, USA}

\author{Liting Xiao}
\affiliation{LIGO Laboratory, California Institute of Technology, MS 100-36, Pasadena, California 91125, USA}

\date{\today}

\begin{abstract}

We describe the methods used to construct the aligned-spin template bank of
gravitational waveforms used by the GstLAL-based inspiral pipeline to analyze
data from the second observing run of Advanced LIGO and the first observing run of Advanced Virgo. The bank expands upon the parameter space covered during Advanced LIGO's first observing run, including coverage for merging compact binary systems with total mass between
2\,$\msun$ and 400\,$\msun$ and mass ratios between 1 and 97.988. Thus the
systems targeted include merging neutron star-neutron star systems, neutron
star-black hole binaries, and black hole-black hole binaries expanding into the intermediate-mass
range. Component masses less than 2\,$\msun$ have allowed (anti-)aligned spins between $\pm0.05$ while
component masses greater than 2\,$\msun$ have allowed (anti-)aligned between
$\pm0.999$. The bank placement technique combines a stochastic method with a
new grid-bank method to better isolate noisy templates, resulting in a total of
677,000 templates.

\end{abstract}

\maketitle

\section{Introduction}\label{sec:intro}

The first observing run (O1) of the advanced Laser Interferometer
Gravitational-wave Observatory (LIGO)~\cite{AdvLIGO1, AdvLIGO2} detectors collected data from September 12, 2015 to January 19, 2016, during which two gravitational-wave (GW) signals
were detected at greater than 5$\sigma$, GW150914~\cite{GW150914} and
GW151226~\cite{GW151226} from the mergers of two binary black hole (BBH)
systems. The second observing run (O2) of Advanced LIGO ran from November 30,
2016 to August 26, 2017, with Advanced Virgo~\cite{AdvVirgo} joining the run for the month of
August. By the end of O2, GWs from a total of ten binary black hole
mergers had been observed across O1 and O2~\cite{GW170104, GW170608, GW170814, O1-O2Catalog}. In addition, the low-latency discovery of GWs from a binary neutron star (BNS) merger~\cite{GW170817} was also reported from O2.

These types of signals are targeted by all-sky, matched-filter-based searches
including GstLAL~\cite{O1Methods, gstlal},
PyCBC~\cite{pycbc1,pycbc2,pycbc3,pycbc4} and MBTA~\cite{MBTA}.
Matched-filter-based searches correlate detector data with waveforms predicted
by general relativity, drawn from a template bank. The template bank contains
waveforms covering a multi-dimensional parameter space of component masses and
spins. If a template closely matches a hidden signal in the data, a peak (or
trigger) in the correlation time series will be produced. The search pipelines
then employ a number of techniques to ensure that triggers are found in
operating detectors within the inter-site propagation time and that the signal
has the expected morphology and amplitude.

Such sources are also targeted by un-modeled searches that do not use template waveforms, for example, the coherent WaveBurst algorithm~\cite{cWB}. Such searches are especially effective for the heavier mass binaries, which have shorter template waveforms within the detector frequency band and hence look similar to short duration noise transients or glitches, making it difficult for them to be recovered by the template based searches. The un-modeled searches requiring minimal assumptions about the waveform, have so far been more sensitive to the heavier binaries~\cite{IMBHO1-O2}. They are also expected to be more sensitive to signals from such heavier binaries including non-fundamental higher order mode effects, than some matched-filter based searches~\cite{IMBHHM}. Also, techniques to construct template banks directly from numerical relativity simulations have also been explored, as described in~\cite{nrbank}. In this paper we will focus on the GstLAL-based inspiral search.  

The GstLAL-based inspiral search pipeline (henceforth referred to as GstLAL) operates in two modes, a low-latency online mode
and an offline deep-search mode. In this paper, we describe both the template bank used in the online mode for issuing low-latency alerts
to astronomy partners as well as the template bank used in the offline mode for the deep analysis of O2 data.

During O1, the matched-filter-based searches of PyCBC and GstLAL used a common
template bank with total masses between 2\,$\msun$ and
100\,$\msun$~\cite{CBCcompanion, O1BBHobs}, to search for stellar-mass binary black holes. For O2, separate banks were constructed and utilized~\cite{O2Pycbc}, to enhance the independence of
search-pipeline results. Additionally, for O1, a separate search for GWs from
intermediate-mass black hole binaries (IMBHBs) was performed using a template
bank with total mass between 50\,$\msun$ and 600\,$\msun$~\cite{O1IMBH}. The
IMBHB search thus partially overlapped the stellar-mass search in the mass range between 50\,$\msun$ and 100\,$\msun$, which resulted in complexities in assigning significances to candidate
events recovered by both the IMBHB and the stellar-mass GstLAL searches with differing sensitivities. For O2, an integrated search was implemented with a template bank covering total mass up to 400\,$\msun$.

The paper is organized in the following way. In Section~\ref{sec:design} we
describe the design and construction of the integrated template bank used by
GstLAL for the analysis of O2 data. In Section~\ref{sec:effectual} we describe
the performance of the bank in recovering simulated signals from a variety of
astrophysical populations. We present our conclusions in
Section~\ref{sec:conclusion}.

\section{Design and construction of the O2 bank} \label{sec:design}

\subsection{Astrophysical source targets} \label{subsec:astro} The O2 GstLAL search targeted GW
signals from merging binary compact objects with component masses between
1\,$\msun$ and 399\,$\msun$. These include binary systems with two neutron
stars (BNS), two black holes (BBH), or a neutron star and a black hole (NSBH).
This component mass region is known to be populated with compact objects
produced from the collapse of massive stars. With stellar evolution models,
neutron stars can form in the mass range between 1\,$\msun$ and
3\,$\msun$~\cite{Rhoades1974, Kalogera1996, OzelNSLower, Lattimer2012,
Kiziltan2013} although there is only one observed neutron star with a mass
larger than 2\,$\msun$~\cite{MassiveNS}, and those in binaries do not approach
2\,$\msun$~\cite{Ozel2016}. Stellar evolution models also predict that black
holes may exist with a minimum mass down to 2\,$\msun$~\cite{Shaughnessy2005}
and a maximum mass up to 100\,$\msun$ or potentially
higher~\cite{Belczynski2014, deMink2015}. Black holes with masses between
$\sim$100\,$\msun$ and $\sim10^5$\,$\msun$ are classified as intermediate-mass
black holes and could have formed through hierarchical merging of lower mass
black holes~\cite{Miller2004}. This search is also sensitive to GWs from
binaries of primordial black holes (PBH), formed from over-dense regions in the
early universe. However, distinguishing a PBH GW signal from a conventional
stellar-evolution black hole GW signal would not be possible with this search
and is instead pursued in a separate search of the sub-solar mass
region~\cite{Subsolar2018}.

We also define different ranges of allowed angular momentum for component
neutron stars and component black holes. We consider the dimensionless
spin parameter $\chi= c\left| \vec{S}\right| /Gm^2$ where $\vec{S}$ is the angular
momentum and $m$ is the component mass. Observations of the fastest spinning
pulsar constrain $\chi \lesssim 0.4$~\cite{Hessels2006} while pulsars in
binaries have $\chi \le 0.04$~\cite{Kramer2009}. X-ray observations of
accreting BHs indicate a broad distribution of BH spins~\cite{Fabian2012,
Gou2011, Mcclintock2011}, while the relativistic Kerr bound $\chi \le 1$ gives
the theoretical constraint~\cite{MisnerGrav}.

\begin{figure}[hbt!]
\includegraphics[width=0.45\textwidth]{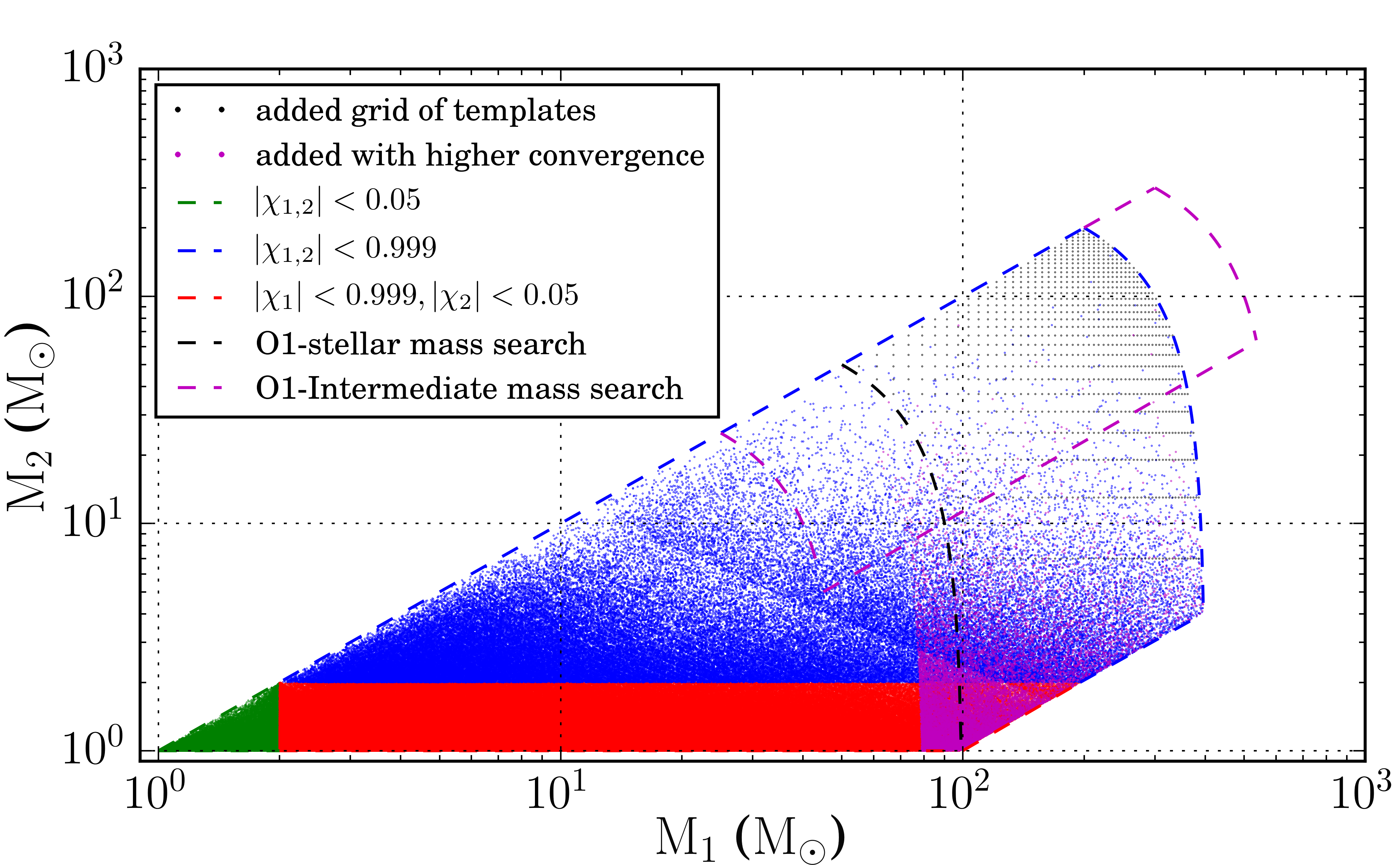}
\caption{(Color online) The template bank used by the O2 GstLAL offline search in component mass space.The templates representing the different astrophysical populations are shown in green for BNS, blue for BBH, and red for NSBH.}
\label{fig:H1L1V1-GSTLAL_INSPIRAL_PLOTBANKS_bank_regions_imri-0-0.png}
\end{figure}

These observations and evolution models inform the ranges of parameters we
define for our template banks. As shown in
Fig.~\ref{fig:H1L1V1-GSTLAL_INSPIRAL_PLOTBANKS_bank_regions_imri-0-0.png}, we can
see the BNS, NSBH, and BBH populations represented in the O2 GstLAL offline search. We impose an additional
constraint on the component dimensionless spins of template waveforms by
requiring their orientations to be aligned or anti-aligned with the orbital
angular momentum of the binary $\hat{L}$. Then the dimensionless projections of
the component spins along $\hat{L}$ are defined as $\chi_i \equiv c\left|
\vec{S}_i \cdot \hat{L} \right| /Gm_i^2$. The region in green marks the BNS
templates with component masses between 1\,$\msun$ and 2\,$\msun$ and
(anti-)aligned dimensionless spin magnitudes with $ \chi_{1,2}<0.05$. This
$\chi$ limit is motivated by the observational limit of $\chi \le 0.04$ but
with some added uncertainty. The region in blue marks the BBH templates with
component masses between 2\,$\msun$ and 399\,$\msun$, mass ratios greater than 1 and less than 98.988, which is the maximum for the bank and (anti-)aligned dimensionless spin magnitudes with $\chi_{1,2} <0.999$. This $\chi$ limit is chosen to be as close to the theoretical limit of 1 as possible with current waveform
approximants, as described in Section~\ref{sec:approximant}. The templates in red mark the NSBH
range with the neutron star mass between 1\,$\msun$ and 2\,$\msun$ and the
black hole mass between 2\,$\msun$ and 200\,$\msun$. For these systems, neutron
stars have $ \chi_{1,2}  <0.05$ and black holes have $\chi_{1,2} <0.999$.

\begin{figure}[hbt!] \includegraphics[width=0.45\textwidth]{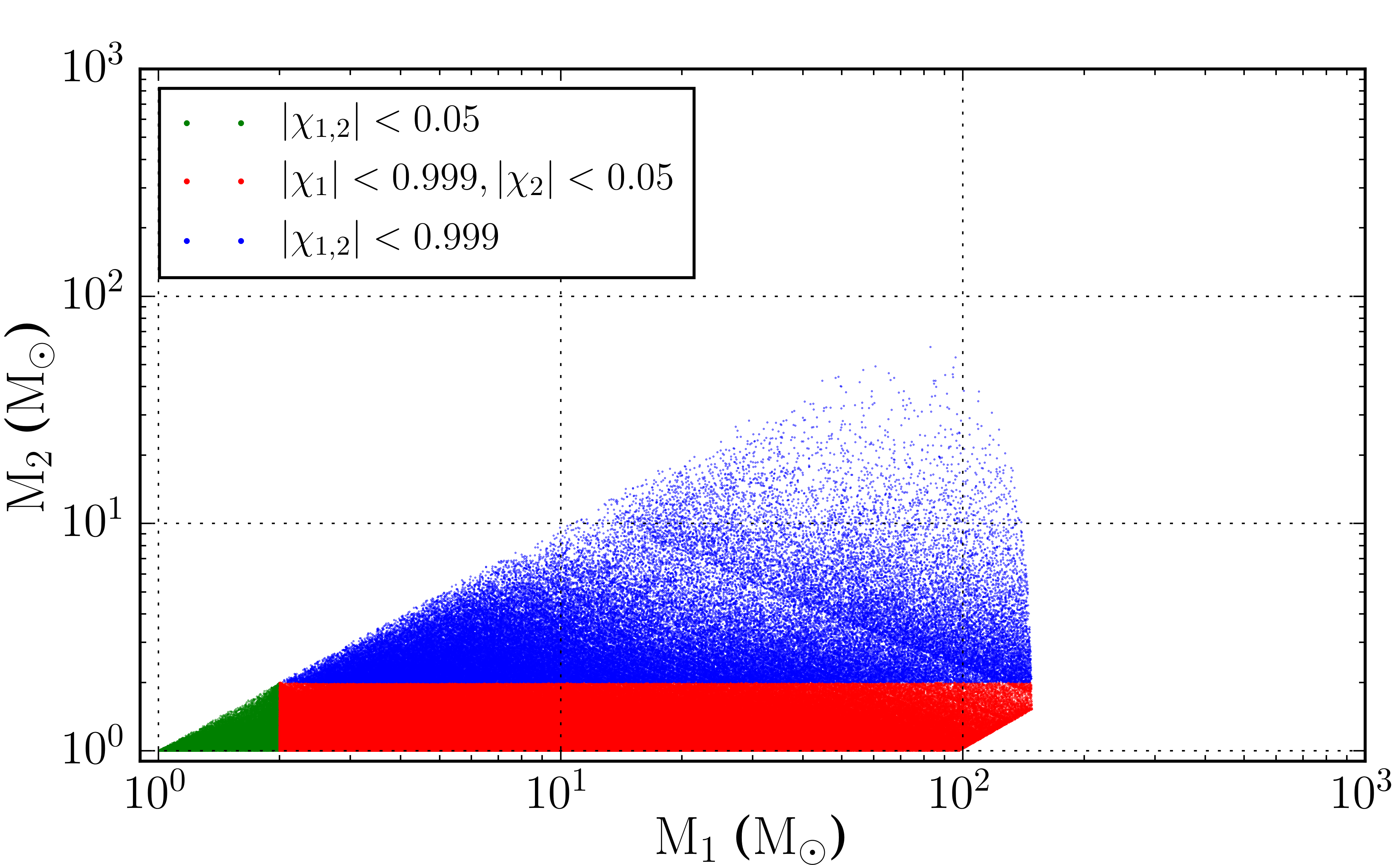}
\caption{\label{online_bank.png}(Color online) The template bank used by the O2 GstLAL online search in component mass space. The templates representing the different astrophysical populations are shown in green for BNS, blue for BBH, and red for NSBH.} \end{figure}

In Fig.~\ref{online_bank.png}, we can see the BNS, NSBH, and BBH populations represented in the O2 GstLAL online search.
The BNS templates cover the same component mass and dimensionless spin magnitude
range as the offline bank. However, an additional restriction
employed in total mass ($M>150\,\msun$) resulted in different component mass ranges for
NSBH and BBH templates for the online template bank. The maximum allowed total
mass is 150\,$\msun$, to remove high-mass templates which correspond to short waveforms
that recover short transient noise fluctuations (glitches) at a high rate.

\subsection{Construction of the O2 bank}

The construction of a template bank relies on a number of parameters, including the selection of
a representative noise power spectral density $S_n(f)$ and appropriate waveform models,
the waveform starting frequency $f_\mathrm{low}$, the placement method, and a specified minimum
fitting factor criteria~\cite{privitera2014improving, fittingfactor, FFdef} for all templates in the bank.

The minimum fitting factor describes the effectualness of a template bank in recovering
astrophysical sources. To define this quantity, we note that the matched filter output is maximized
when a template waveform exactly overlaps the signal waveform. This optimization is impossible in practice,
however, since the template bank samples the parameter space discretely while
astrophysical sources arise from a continuum. Regardless, it is useful to
quantify the degree to which two waveforms, $h_1$ and $h_2$, overlap. The
overlap is defined as the noise-weighted inner product integral~\cite{privitera2014improving}:
\begin{align} \label{eq:overlap}
(h_{1} | h_{2}) = 2 \int^\infty_{f_\mathrm{low}} \frac{\tilde{h}_1(f)\tilde{h}^{*}_2(f) + \tilde{h}_1^{*}(f)\tilde{h}_2(f)}{S_n(f)}df,
\end{align}
where $f_\mathrm{low}$ was set to 15\,Hz, as motivated by the noise power spectral density described in
Section~\ref{sec:psd}.

The \emph{match} between two waveforms is then defined as the noise-weighted inner product~\cite{privitera2014improving, fittingfactor}, maximized over a set of parameters denoted by $\phi$. For precessing signals, this overlap calculation considers only the (2,2) mode and maximizes over the template's coalescence phase, polarization and sky position while the overlap calculation for the higher order mode waveforms maximizes over only the polarization and sky position.

\begin{align} \label{eq:Match} M(h_{1}, h_{2}) = \underset{\phi}{\text{max}} (h_{1}|h_{2}(\phi)) \end{align}
This defines the percent of signal-to-noise ratio (SNR) retained when recovering waveform $h_2$
with the (non-identical) waveform $h_1$. Then, the fitting factor is the related quantity used
in describing the effectualness of template banks:
\begin{align} \label{eq:FF} FF(h_{s}) = \underset{h \in \{h_{b}\}}{\text{max}} M(h_s, h) \end{align}
where $h_b$ is the set of templates in the bank and $h_s$ is a signal waveform
with parameters drawn from the continuum.  For the aligned spin waveforms, the $FF$ is calculated by maximizing the noise-weighted inner product only over the templates. The fitting factor describes the fraction of SNR retained for arbitrary signals in the parameter space covered by the bank. Typically,
compact binary coalescence searches have required a fitting factor of $97\%$ to ensure
that no more than $\sim10 \%$ of possible astrophysical signals are lost due
to the discrete nature of the bank. As described in Sect.~\ref{sec:placement}, we
use a hierarchical set of fitting factor requirements to construct the bank.

\subsubsection{Modeling the detector noise}\label{sec:psd}
\begin{figure}[hbt!]
\includegraphics[width=0.4\textwidth]{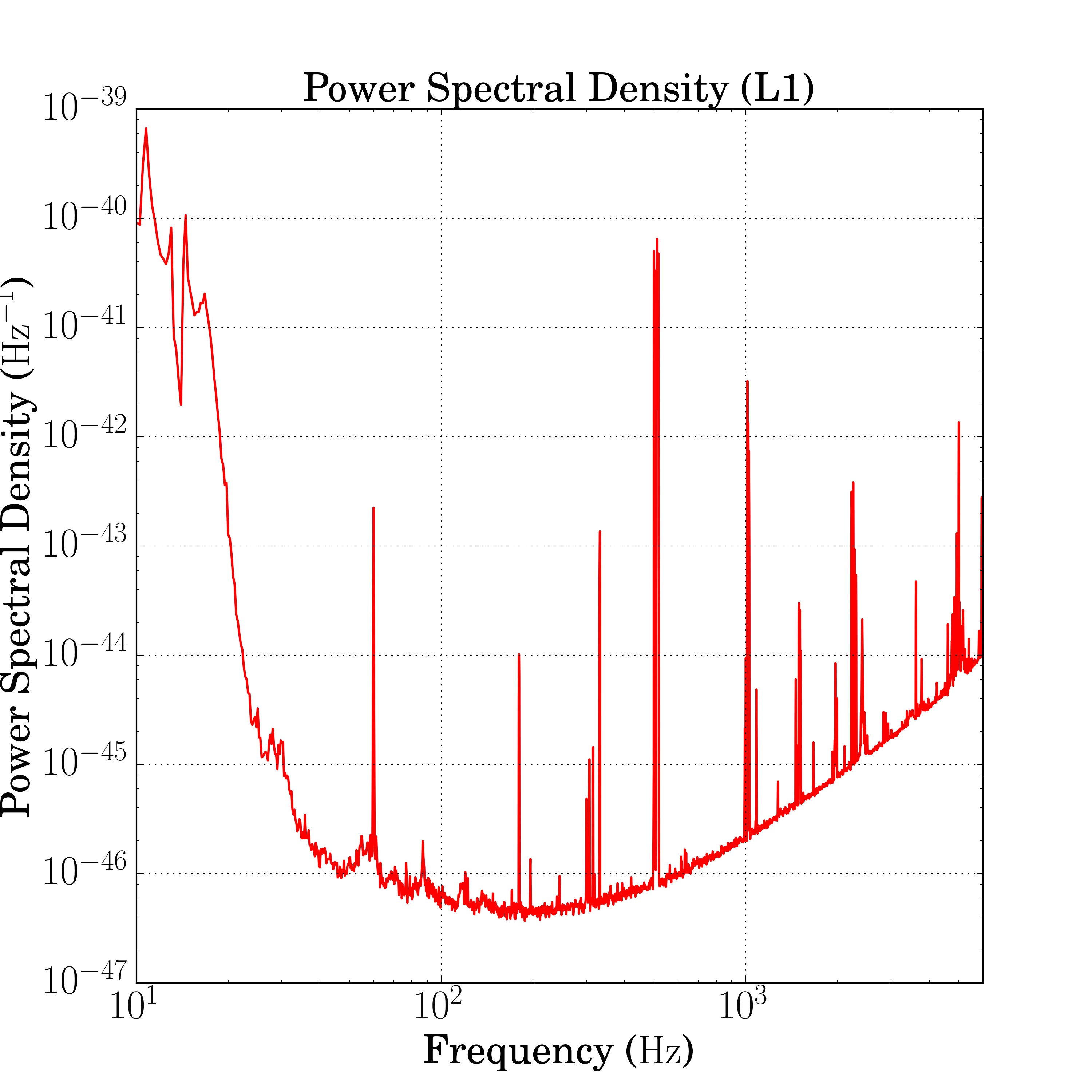}
\caption{\label{psd.pdf}(Color online) Representation of the model power spectral density of
detector noise. This was used to construct the O2 template bank. } \end{figure}

The noise power spectral density (PSD) as shown in Fig.~\ref{psd.pdf} was used to
compute the overlap integrals in the construction of the O2 template bank.
This projected O2 sensitivity curve was produced by combining some of the
best LIGO L1 sensitivities achieved before the start of O2. At low frequencies,
below 100\,Hz, the best sensitivity is taken from L1 measurements during commissioning
in February 2016. At high frequencies, above 100\,Hz, the best sensitivity is taken from
L1 during O1, with projected improved shot noise due to slightly higher input power and
improved efficiency of the readout chain. Calculation of
this PSD has been documented in~\cite{psd}.

\subsubsection{Waveform approximants} \label{sec:approximant}
 Gravitational waveforms from compact binary mergers are described by
 17 intrinsic and extrinsic parameters. However, as demonstrated in~\cite{ajith2014effectual}, for template placement purposes, we can parameterize these systems by three parameters composed of
component masses $m_i$ and a reduced-spin parameter as defined in~\cite{ajith2014effectual}, comprising the non-precessing spin of the waveform, which is a function of
the dimensionless spin parameters $\chi_i$ for $i=1,2$.

Above a total mass of 4\,$\msun$, the waveforms of the binary systems are
computed using the effective-one-body formalism (SEOBNRv4\textunderscore
ROM)~\cite{SEOBNRv4ROM}, combining results from the post-Newtonian approach,
black hole perturbation theory and numerical relativity to model the complete
inspiral, merger and ringdown waveform. For binaries with total mass $\leq 4 \msun$, waveforms are approximated by post-Newtonian inspiral templates accurate to third-and-a-half order (the TaylorF2 approximant)~\cite{TaylorF2,CBCcompanion}. The extent of the present parameter space
covered by the template bank is limited by the availability of waveform models
and the sensitivity of the present search. We neglect the effect of precession
and higher order modes in our templates.

\subsubsection{Template placement}\label{sec:placement}
Both the O2 offline and online template banks were created in the same way, by constructing two sub-banks that were added together. For systems with total mass $2\msun \le M \le 4\msun$ where the TaylorF2 approximant is used, the templates were first laid out using a
geometric metric technique~\cite{geometric}. This geometric bank was used as a coarse seed bank for an additional stochastic method placement~\cite{ajith2014effectual, stochastic} where templates were laid down densely enough to allow only a 3\% loss in SNR from a template not exactly matching a gravitational-wave signal, thus satisfying a minimum match set to 97$\%$. For systems with total mass greater than 4\,$\msun$, where the SEOBNRv4\textunderscore ROM approximant is used, a coarse bank was first generated with the stochastic method but
with a very low minimum match. Again this stochastic bank was used as a coarse seed bank for an additional stochastic method placement with a minimum match set to 97$\%$. Additionally, only waveforms with a duration longer than a threshold of 
0.2\,s chosen in an ad hoc manner, were retained initially, to avoid recovering short transient noise glitches, which are mostly seen to ring up heavier mass templates with shorter duration. The two sub-banks were added to form the full bank with a total of 661,335 templates.

\begin{figure}[hbt!]
\includegraphics[width=0.45\textwidth]{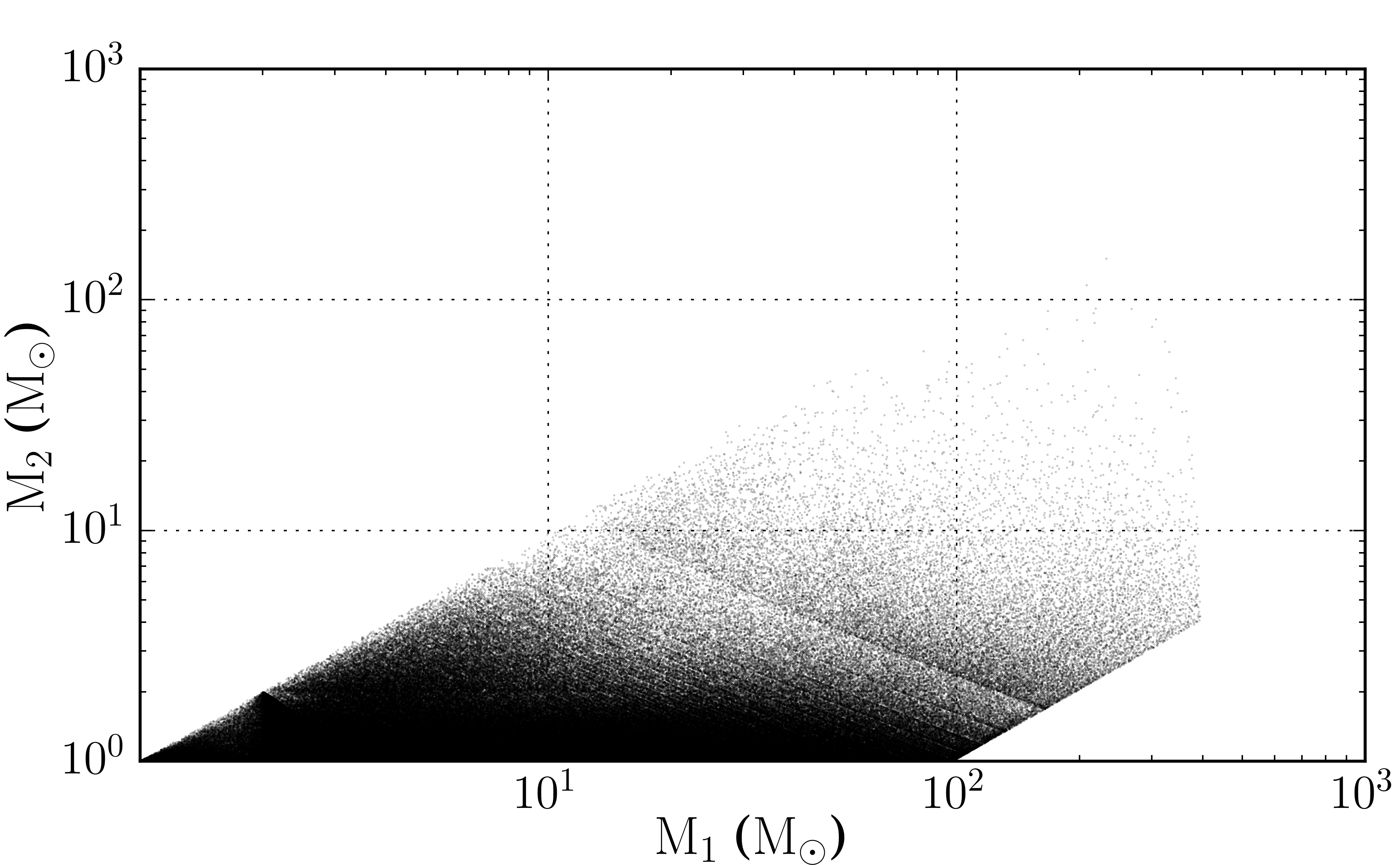}
\caption{\label{gstlal_bank_m1m2.png}A visual representation of the original O2
bank in the component mass space, containing a total of 661,335 templates
placed with a minimal match of 97$\%$.} \end{figure}

The original O2 offline bank, as shown in Fig.~\ref{gstlal_bank_m1m2.png} aided in the discovery of
one of the earliest events detected during O2, GW170104 [3]. The higher density of
the bank at lower masses is expected because low mass systems have
longer waveforms and spend more time in the detectors' sensitive frequency band. This
enables the matched-filter search to better distinguish between two different
low mass systems. This also means that more templates are required in the lower mass region of the bank for
the required minimum match. At the highest masses, the waveforms contain very few
cycles and very few templates are required for coverage in this region.

Early in O2, short duration glitches were found to be particularly problematic for the online search, even with a duration cut of 0.2\,s applied. Thus, to avoid delays in delivering low-latency gravitational-wave triggers, only waveforms with a total mass $<150\,\msun$ were retained in the online bank, based on our observations of the heavier mass templates being the most likely to recover noise triggers. This online bank, as shown in Fig.~\ref{online_bank.png}, was used for the entirety of the O2 observing run.

\subsubsection{Additional coverage in the offline bank}\label{sec:overcoverage}
As outlined in Section~\ref{sec:implementation}, templates are grouped into bins by the GstLAL search such that all of the templates in any given bin have similar responses to noise, i.e. the distribution of SNR and $\chi^2$ in noise are similar for all of the templates in each bin~\cite{O1Methods}. It was uncovered partway through O2 that the lower density of templates in the high mass part of the offline bank (total mass $>$ 80 $\msun$) was resulting in templates with very different background noise properties to be grouped together. This led to incorrect averaging of noise properties in the high mass groupings of templates and, in turn, resulted in incorrect estimation of the significance of loud coincident noise in time-shifted data from the two detectors ~\cite{O2Methods}. This was
not an issue in the online bank due to the cut at total mass $>150\,\msun$.

Two different re-courses were taken. The offline bank was overpopulated with extra templates in the higher mass region as outlined below. Additionally, the templates in this part of the bank were grouped differently from those in the denser lower mass region. Both of these steps were meant to ensure that more templates with a similar response to background noise can be grouped together, leading to a better estimation of the background noise captured by these templates. The bank was seen to reach its optimal sensitivity when these changes were applied to the higher mass region $80\msun \le M \le 400\msun$. Details of the template grouping methods are given in Ref.~\cite{O2Methods}.

\begin{figure}[hbt!]
\includegraphics[width=0.45\textwidth]{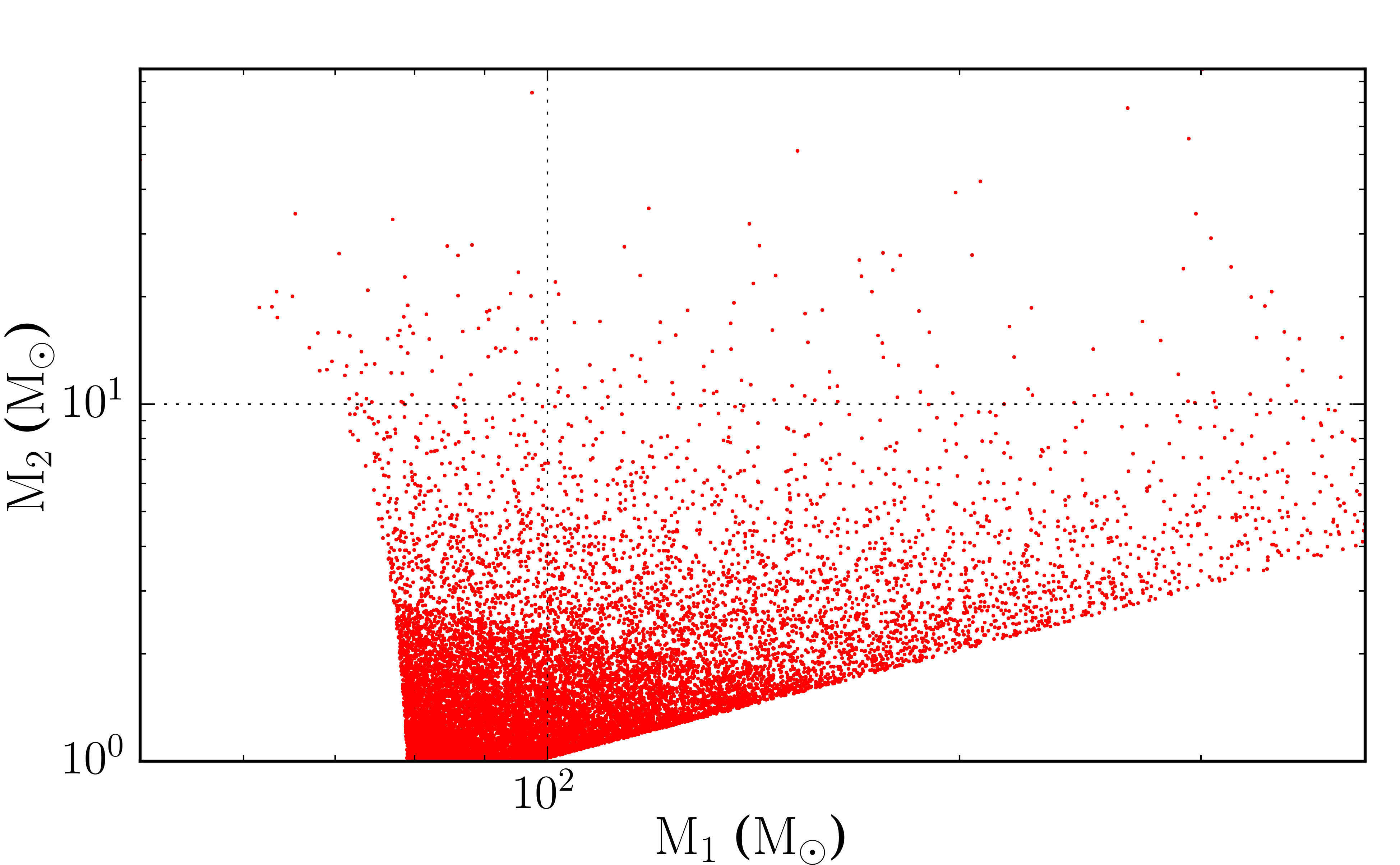}
\caption{\label{mtotalcut80_m1m2.png}(Color online) The bank of extra 14,665 templates that
were added to the initial O2 bank, with a 98$\%$ minimum match above a total
mass of 80 $\msun$ in the component mass space.} \end{figure}

Regarding the additional coverage, extra templates were added to the initial offline bank in the total mass range of
$80\msun \le M \le 400\msun$ using two methods. As a first step, the original offline bank was used as a seed
for an additional stochastic placement in the total mass range $80\msun \le M \le 400\msun$ with an
increased minimum match of 98$\%$. Initially the template duration threshold of 0.2\,s was chosen in an ad hoc manner in the hope that excluding these short duration templates would reduce the recovery of similar short duration glitches by them. We found this was not the case. Hence additionally, no template duration threshold was used in these extra templates that were added to the mass range $80\msun \le M \le 400\msun$ so as to not exclude the short waveforms corresponding to the heavier mass systems. In the region $M \le 80\msun $, the duration of the templates being much longer than 0.2\,s, they remained unaffected by this duration threshold. A total of 14,665 templates, as shown in Fig.~\ref{mtotalcut80_m1m2.png}, were added to the initial offline bank.

\begin{figure}[hbt!]
\includegraphics[width=0.45\textwidth]{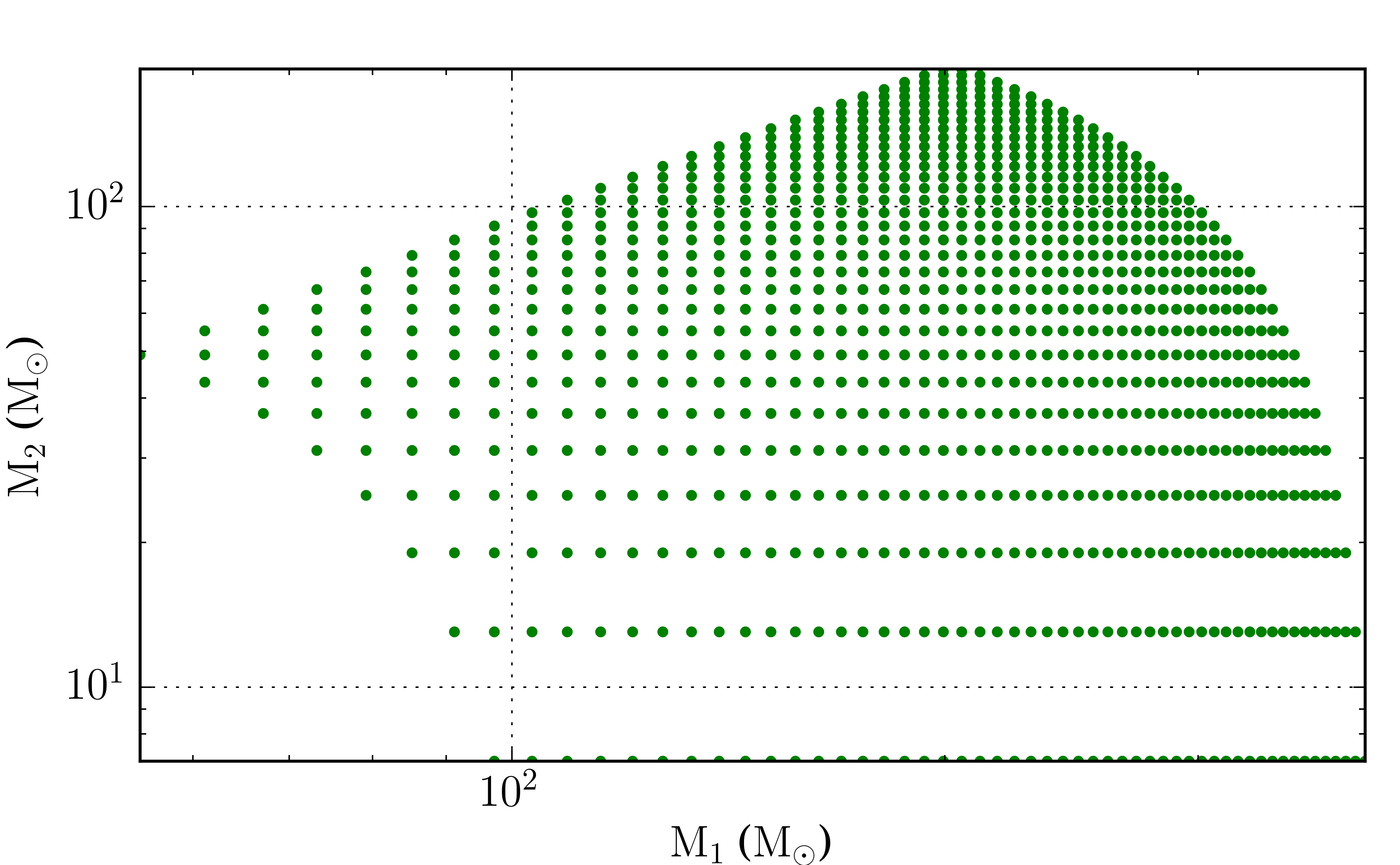}
\caption{\label{uniformgrid_m1m2.png}(Color online) The uniform grid bank with a 1000
templates spanning 100-400 $\msun$ in total mass .} \end{figure}

Despite the increased minimum match, the high mass region of the bank remained sparsely
populated, as the overlap between high mass waveforms with few cycles are generally high. Thus, the minimum match required is already met, without the placement of additional templates. However, short duration glitches are also
recovered by a relatively few number of high mass templates, and if these few glitchy templates are
grouped together for background estimation with quieter templates, they can spoil the sensitivity over a
broad mass range. Thus, we chose to force the placement of additional templates at higher mass using a
uniform grid placement in component mass space for the total mass range between $100\msun \le M \le 400\msun$, with mass ratios between 1 and 97.988.
A total of 1000 templates were placed without any limitations on the waveform duration. This gridded bank, as
shown in Fig.~\ref{uniformgrid_m1m2.png}, was then added to the offline bank produced in the previous step.

All together, the final, improved O2 bank has a total of about 677,000 templates, as shown in
Fig.~\ref{fig:H1L1V1-GSTLAL_INSPIRAL_PLOTBANKS_bank_regions_imri-0-0.png}.

\subsection{Implementation in the GstLAL pipeline}\label{sec:implementation}
The GstLAL-based inspiral search is a matched-filtering pipeline. The noise-weighted inner
product of each whitened template with the whitened data produces the signal-to-noise
ratio (SNR). Both signals and glitches can produce high SNR, thus a number of additional
consistency and coincidence checks are implemented in the pipeline, as detailed in Ref.~\cite{O2Methods}.
In order to access the full waveform of binary systems up to 400\,$\msun$ that merge at lower frequencies,
the filtering frequency was reduced from 30\,Hz in the O1 search to 15\,Hz.

In the GstLAL pipeline, the background characteristics are estimated for each detector independently using non-coincident triggers that occurred during times they could have formed a coincidence, i.e. non-coincident triggers that occurred when more than one detector had science quality data. This is done for several different bins of templates across the template bank, thus each bin collects its own background statistics used for assigning likelihood-ratios to candidates in that bin. In the region of the template bank where the total mass is less than 80\,$\msun$, the templates are binned first in the effective spin and then in chirp mass (as defined below); in the high mass region, templates are binned by template duration. These two techniques were experimentally found to group templates together that have similar responses to noise. More information on the spin and chirp mass binning can be found in Ref.~\cite{O1Methods}, and more information on the duration binning can be found in Ref.\cite{O2Methods}. For the purpose of background estimation, templates are grouped together so that each group has templates with similar response to background noise.
Noise properties are averaged separately for each group. Before O2, templates were grouped according to two composite parameters that characterize the waveform inspiral to leading order. As also pointed out in~\cite{O1Methods, O2Methods}, these were the chirp mass of the binary $\mchirp$ and the effective spin parameter $\chi_\mathrm{eff}$. The chirp mass is 
\begin{align}
\Mc &= \frac{(m_1 m_2)^{3/5}}{(m_1 + m_2)^{1/5}}.  \end{align} The effective
spin parameter is defined as \begin{align} \chi_\mathrm{eff} &\equiv \frac{m_1
\chi_1 + m_2 \chi_2}{m_1 + m_2}, \end{align} and acts as a mass-weighted
combination of the spin components (anti-)aligned with the total angular
momentum.

However, as described in Sect.~\ref{sec:overcoverage}, extra templates were placed in the
high mass region of the offline bank, to better capture the properties of the noise in that regime.
Templates above a total mass of 80\,$\msun$ were then grouped by template duration from 15\,Hz
rather than the $\mchirp$ and $\chi_\mathrm{eff}$ binning used at lower masses. Template duration
better characterizes the waveform merger and ringdown, the detectable part of the signal
for high mass systems.

\section{Effectualness} \label{sec:effectual}

To assess the effectualness of this template bank, we compute again the
$FF(h_{s})$ as defined in Eq.~\ref{eq:FF} for a collection of simulated signals with
parameters drawn randomly from the covered mass and spin space.  The $FF$ depends on the parameters of the simulations and varies across the parameter space of the bank, hence it can be represented and plotted as a function of two of the parameters of the simulations.  In order to do so, the $FF$  is binned in the two parameters and the mean $FF$  in each bin is plotted~\cite{fittingfactor}. We show such plots in parameters like chirp masses, mass ratios and effective spins. 
\begin{equation} \label{eq:FFav} FF_\mathrm{mean} = \langle FF \rangle \end{equation}

We selected simulated signals from various populations of BNS, NSBH, BBH, and
IMBHB systems to check the effectualness of the bank. The
details of the simulation sets are summarized in
table~\ref{t:banksim_injections}. The simulated signals were chosen to be uniformly distributed on the sky and placed at a fixed luminosity distance of 1 Mpc. Precessing and higher-order mode signals were given binary inclination angles $\iota$ uniformly distributed in $\arccos{\iota}$. The non-spinning IMBHB population described by the EOBNRv2HM~\cite{EOBNRv2HM} waveforms, was also selected to be distributed uniformly in total mass. For the aligned spin binary systems, the extrinsic parameters are not explicitly used in the calculation of effectuality and the waveforms can be generated using the same fiducial extrinsic parameters and are normalized so that the fitting factor of a waveform that has the same parameters as one of the templates is 1. For the precessing simulations, the $FF$ calculation maximizes over coalescence phase, polarization and sky position and for those including higher order mode contributions, it maximizes over only the polarization and sky position. For each signal population, $10^{4}$ simulations were performed.

\begin{table*}[t]
 \centering
 \begin{tabular}{  lllll }
\hline
\hline
Population & Mass($\msun$) & Spin & & Waveform approximant\\
\hline 

BNS & $m_{1,2}  \in  [1,3]$ & $\chi_{1,2} \in [-0.05,0.05]$, aligned & & TaylorF2~\cite{TaylorF2}  \\

BNS & $m_{1,2}  \in [1,3]$ & $\chi_{1,2}  \in [-0.4,0.4]$, precessing & & SpinTaylorT4~\cite{SpinTaylorT4} \\

\multirow{2}*{NSBH} & $m_{1}  \in  [1,3]$    & $\chi_{1} \in  [-0.4,0.4]$, aligned & & SEOBNRv4\_ROM~\cite{SEOBNRv4ROM} \\
 & $m_{2}  \in  [3,97]$ & $\chi_{2}  \in  [-0.989,0.989]$, aligned & & \\
 
 \multirow{2}*{NSBH} & $m_{1}  \in  [3,15]$    & $\chi_{1} \in  [-0.9,0.9]$, precessing & & IMRPhenomPv2~\cite{IMRPhenomPv2} \\
 & $m_{2}  \in  [1,3]$ & $\chi_{2}  \in  [-0.05,0.05]$, precessing & & \\

BBH & $m_{1,2}  \in [2,99]$ & $\chi_{1,2}  \in  [-0.99,0.99]$ aligned & & SEOBNRv4\_ROM~\cite{SEOBNRv4ROM}  \\

BBH & $m_{1,2}  \in [2,99]$ & $\chi_{1,2}  \in  [-0.99,0.99]$ precessing & & SEOBNRv2\_ROM\_DoubleSpin~\cite{SEOBNR_Double} \\

IMBHB & $m_{1,2}  \in  [1,399]$ & $\chi_{1,2}  \in  [-0.998,0.998]$ aligned & & SEOBNRv4\_ROM~\cite{SEOBNRv4ROM} \\

IMBHB & $m_{1,2}  \in  [50,350]$ & Non-spinning & & EOBNRv2HM~\cite{EOBNRv2HM}  \\

\end{tabular}
\caption{\label{t:banksim_injections}Description of different categories of
astrophysical populations, from which random mass and spin parameters were drawn and used to generate
waveforms to check the effectualness of the template bank. Multiple simulation sets of the same population
were used, varying in the type of waveform, mass ranges covered and whether the
spin is aligned to the orbital angular momentum.} \end{table*}

\begin{figure*}[t] 
\centering
\begin{minipage}[b]{0.47\textwidth}
\centering
  \includegraphics[width=\textwidth]{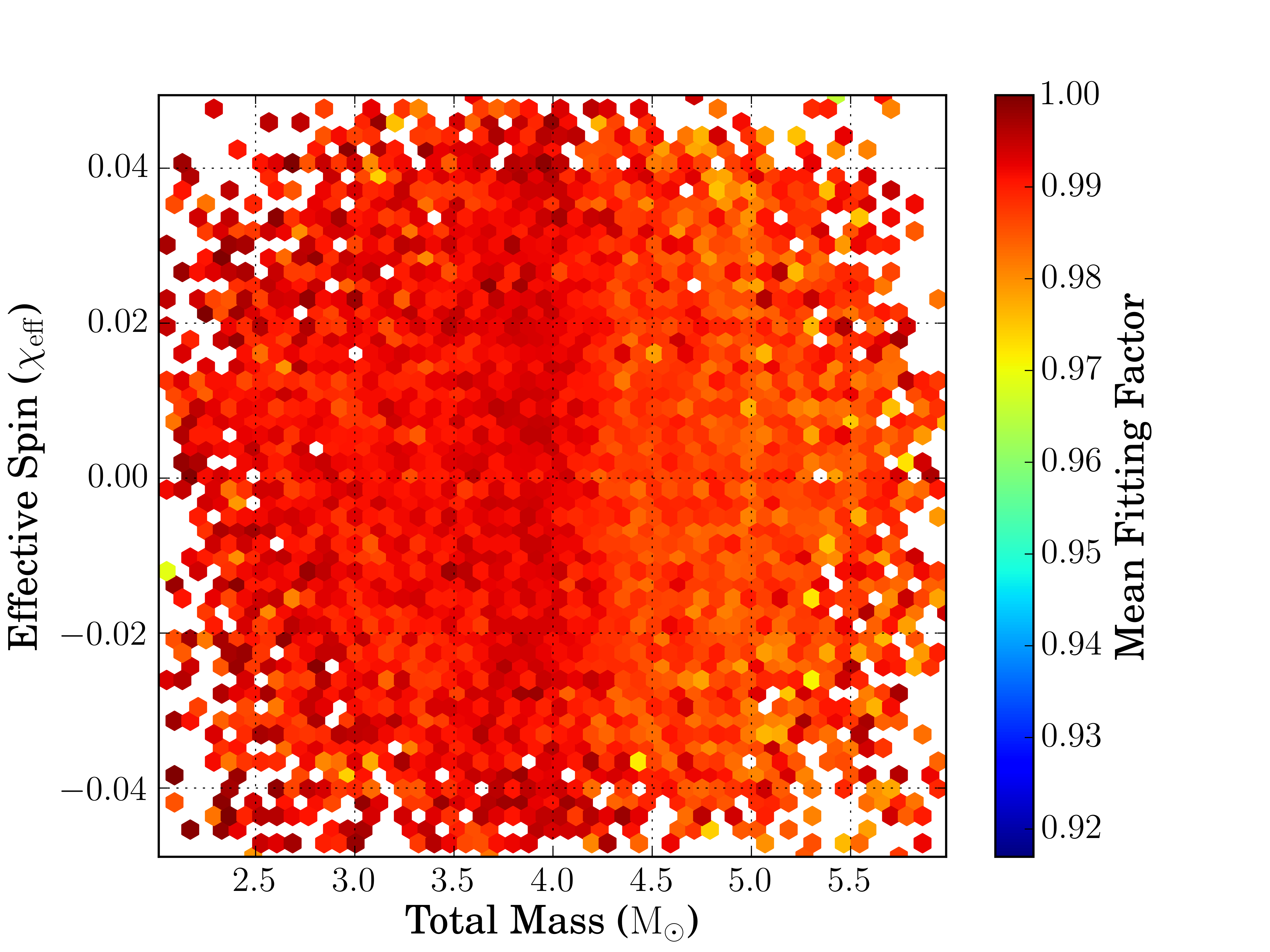}
  \end{minipage}
  \begin{minipage}[b]{0.47\textwidth}
  \centering
  \includegraphics[width=\textwidth]{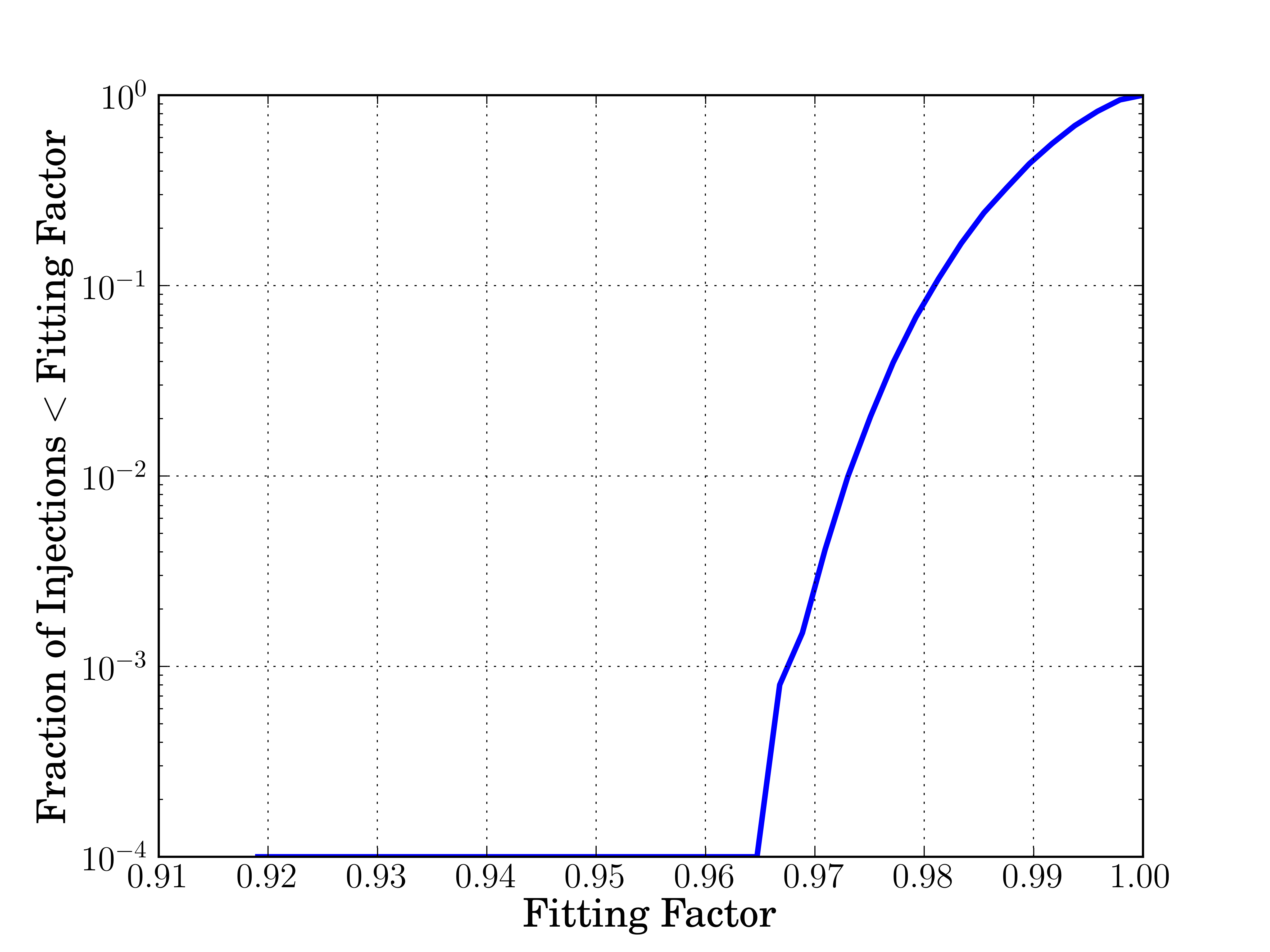}
  \end{minipage}
  \begin{minipage}[b]{0.47\textwidth}
  \centering
  \includegraphics[width=\textwidth]{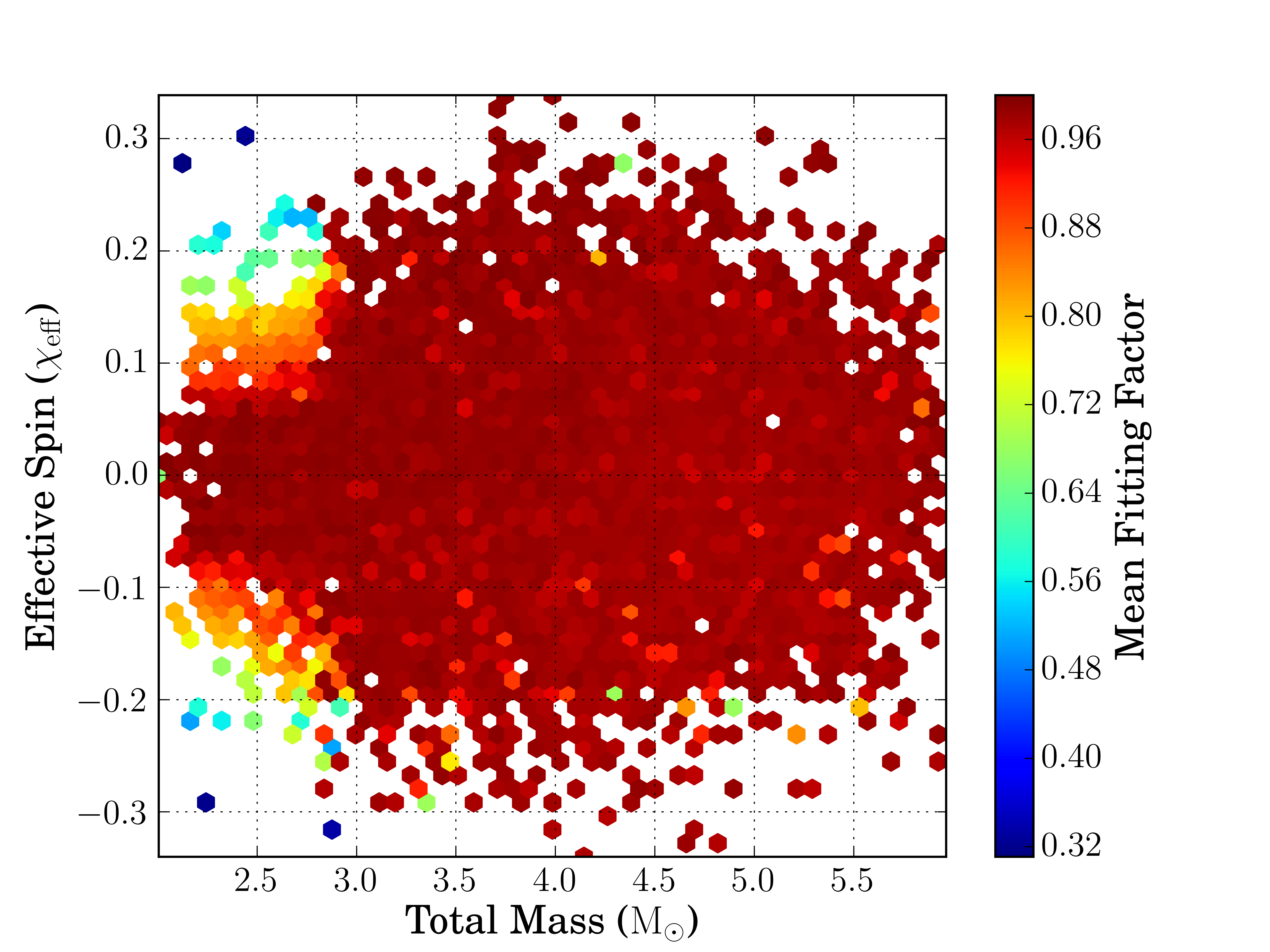}
  \end{minipage}
  \begin{minipage}[b]{0.47\textwidth}
  \centering
  \includegraphics[width=\textwidth]{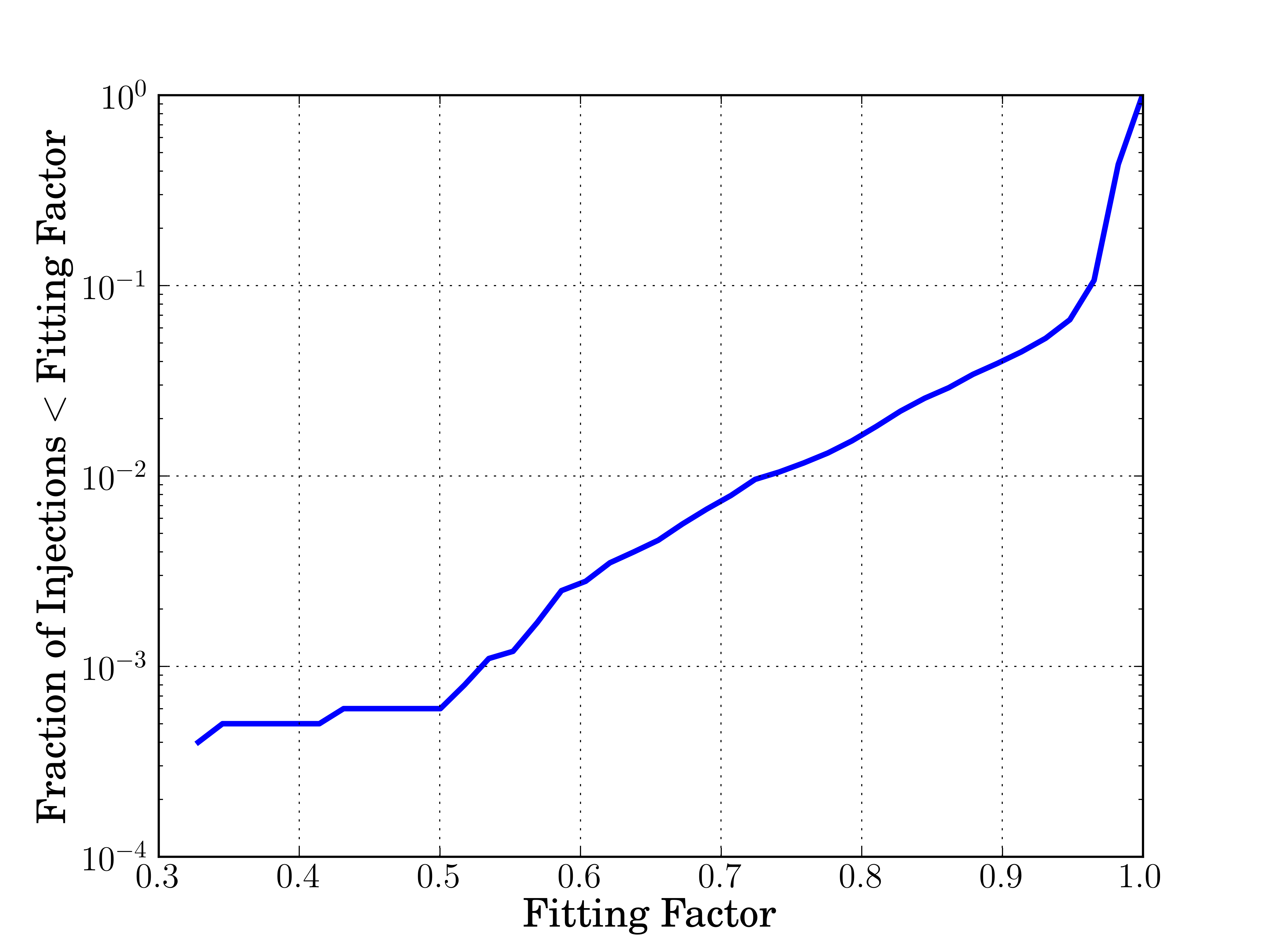}
  \end{minipage}
\caption{\label{fig:bnsff}(Color online) Fitting factors in $M$-$\chi_\mathrm{eff}$ plane for BNS aligned-spin TaylorF2 waveform approximants~\cite{TaylorF2} ({\it top left}) and precessing-spin SpinTaylorT4 waveform approximants~\cite{SpinTaylorT4} ({\it bottom left}). We also include cumulative histograms of the fitting factors of the respective waveforms ({\it top and bottom right}). For the aligned-spin BNS systems, we recover 99.7\% of the injected simulations with fitting factors above 0.97. Hence the majority of fitting factors are above 0.97, except along the low-mass edge of the bank at $M=2.0$ where the fitting factor starts to fall off. The bank is constructed with TaylorF2 waveforms so fitting factors are expected to be at least as high as the required fitting factor of 0.97 to ensure that no more than $\sim10 \%$ of possible astrophysical signals are lost due to the discrete nature of the bank. For the precessing BNS systems we recover 86.1\% of the injected simulations with fitting factors above 0.97, although sensitivity falls off rapidly outside $-0.05 < \chi_\mathrm{eff} < 0.05$ for systems with NS component mass less than 2 \msun. There are no templates placed in this region so the fall off in fitting factor is expected. This also demonstrates that a search based on an aligned-spin template bank can recover precessing-spin signals.}
\end{figure*}

In Fig.~\ref{fig:bnsff}, we can see the fitting factors in the $M$-$\chi_\mathrm{eff}$ plane for BNS aligned-spin TaylorF2 waveform approximants~\cite{TaylorF2} and precessing-spin SpinTaylorT4 waveform approximants~\cite{SpinTaylorT4}. We also include cumulative histograms of the fitting factors of the respective waveforms. 99.7\% of aligned-spin BNS simulations were recovered with a fitting factor above 0.97, while 86.1\% of the precessing BNS simulations were recovered with fitting factors above 0.97. Hence for the aligned-spin BNS systems, the majority of fitting factors are above 0.97, except along the low-mass edge of the bank at $M=2.0\,\msun$ below which no templates are placed. The bank is constructed with aligned-spin TaylorF2 waveforms in this low mass region so fitting factors are expected to be at least as high as the required fitting factor of 0.97 to ensure that no more than $\sim10 \%$ of possible astrophysical signals are lost due to the discrete nature of the bank. For the precessing-spin SpinTaylorT4 waveform approximants the fit falls off rapidly outside $-0.05 < \chi_\mathrm{eff} < 0.05$ for systems with NS component mass less than 2 \msun. There are no templates placed in this region so the fall off in fitting factor is expected. This also demonstrates that a search based on an aligned-spin template bank can recover precessing-spin signals.

\begin{figure*}[t]
\centering
\begin{minipage}[b]{0.47\textwidth}
\centering
  \includegraphics[width=\textwidth]{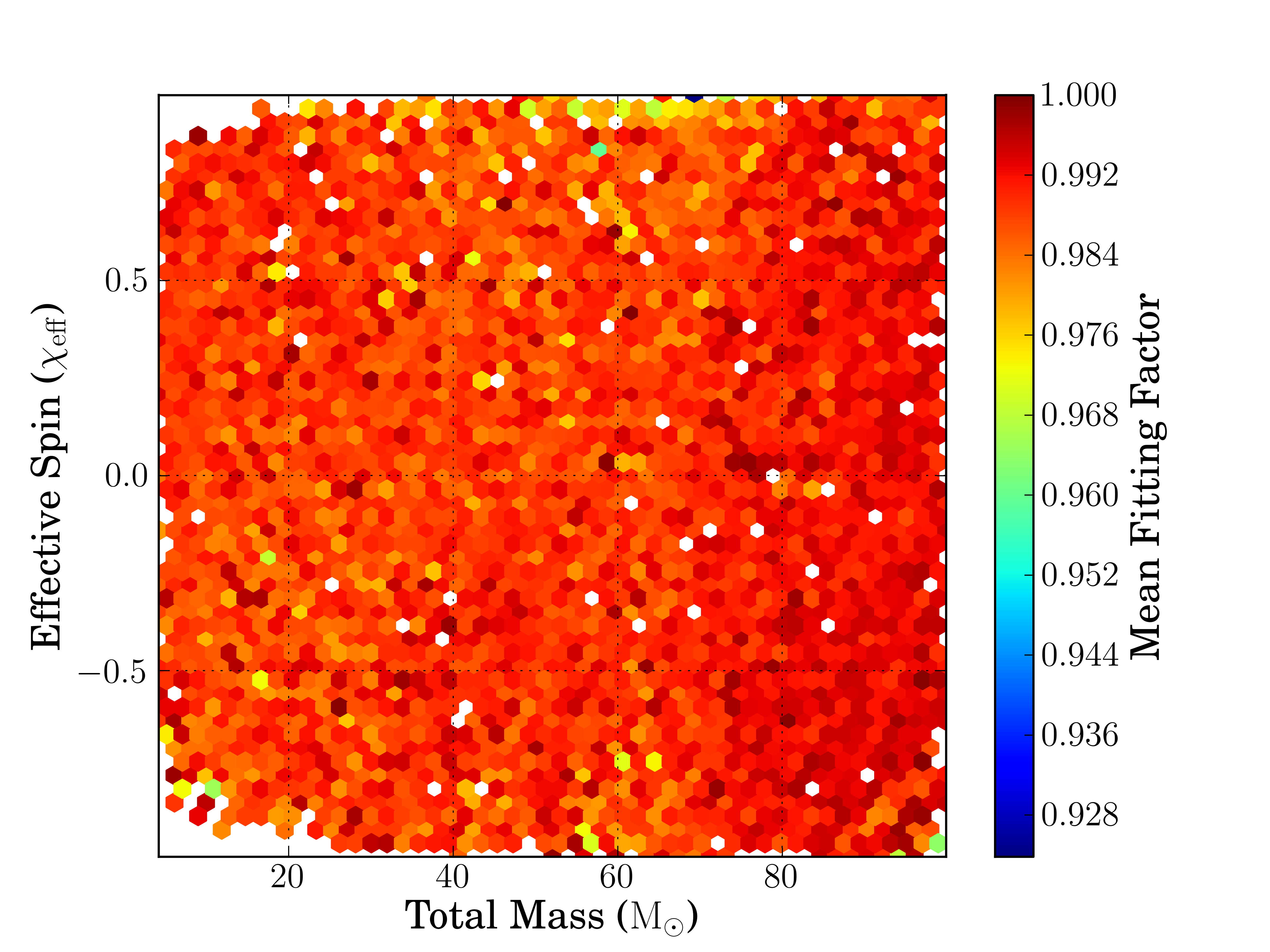}
  \end{minipage}
  \begin{minipage}[b]{0.47\textwidth}
  \centering
  \includegraphics[width=\textwidth]{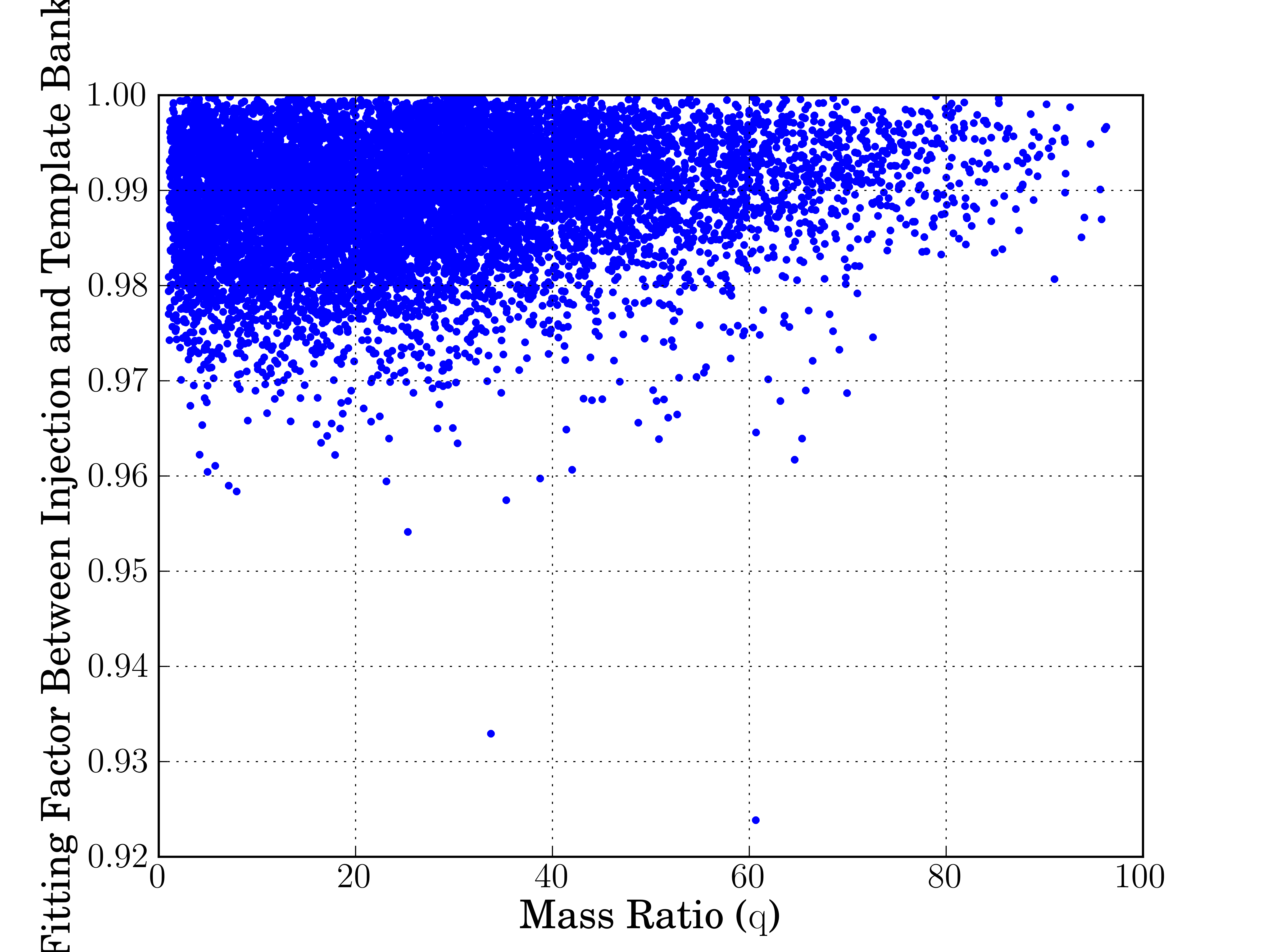}
  \end{minipage}
  \begin{minipage}[b]{0.47\textwidth}
  \centering
  \includegraphics[width=\textwidth]{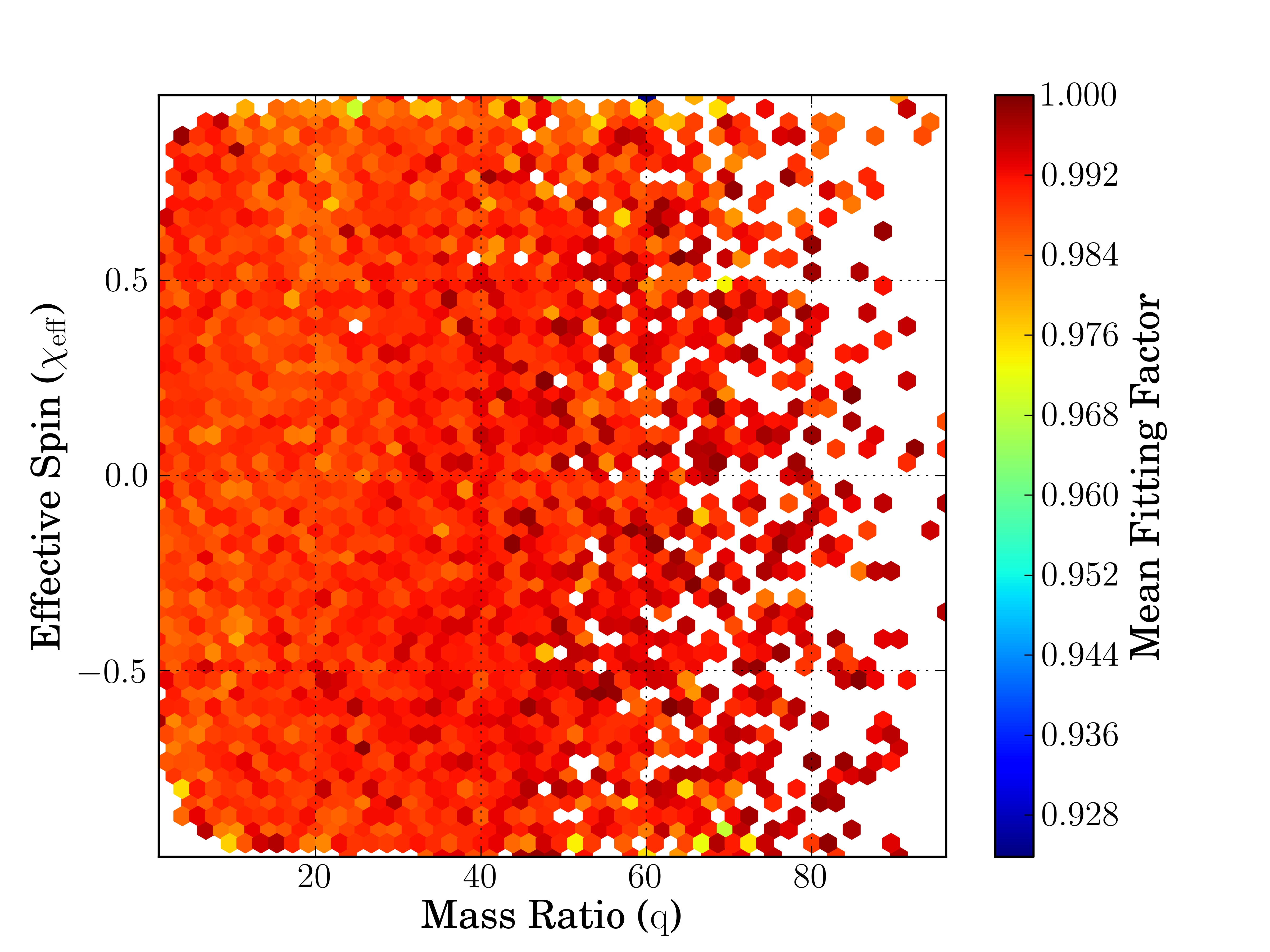}
  \end{minipage}
  \begin{minipage}[b]{0.47\textwidth}
  \centering
  \includegraphics[width=\textwidth]{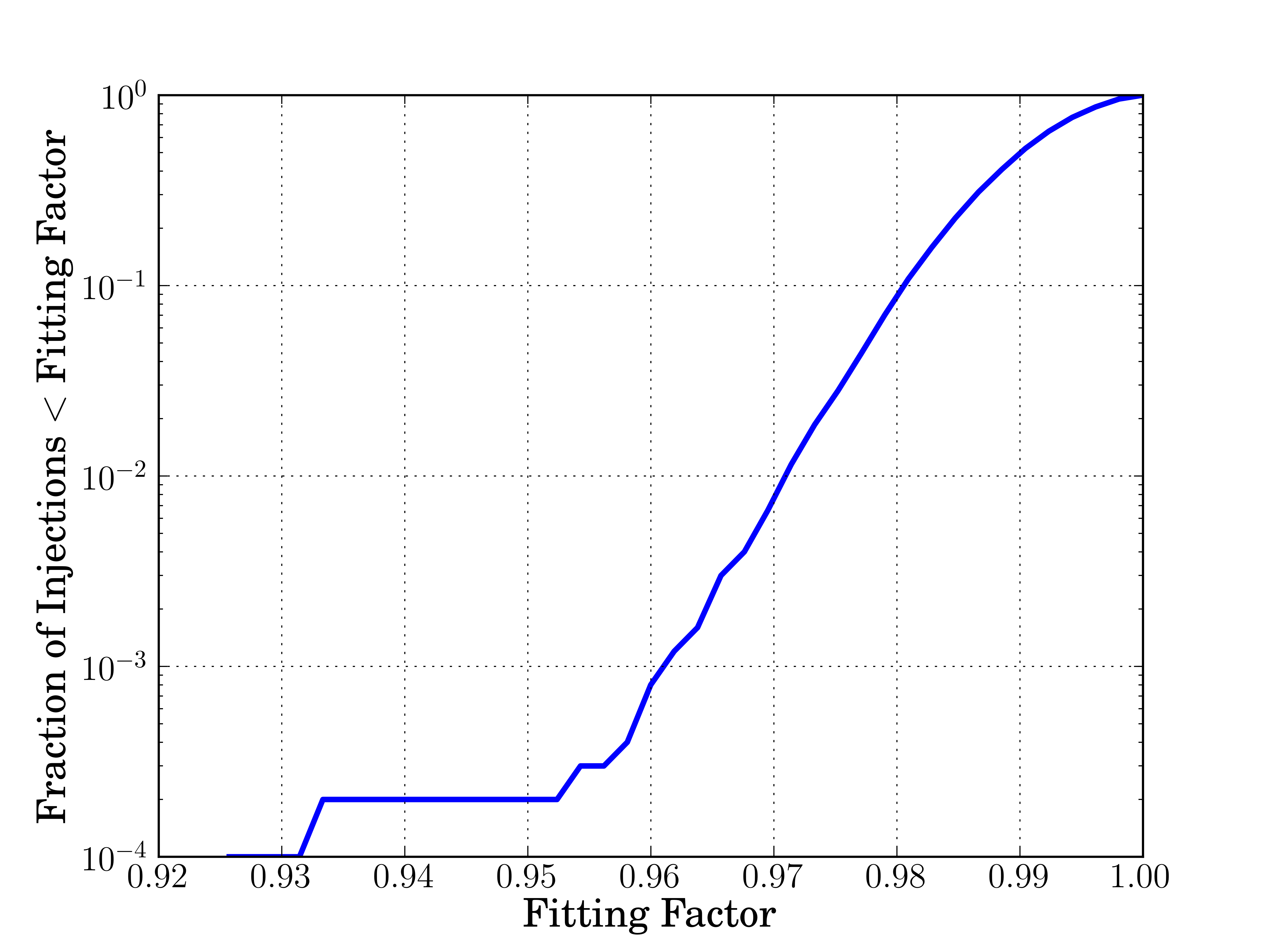}
  \end{minipage}
\caption{\label{fig:nsbhff}(Color online) Fitting factors in $M$-$\chi_\mathrm{eff}$ plane ({\it top left}), as a function of mass ratio (q) ({\it top right}) and in the $q$-$\chi_\mathrm{eff}$ plane ({\it bottom left}) for NSBH aligned-spin SEOBNRv4\_ROM waveform approximants~\cite{SEOBNRv4ROM}. We also include a cumulative histogram of the fitting factors ({\it bottom right}). The majority (99.2\%) of fitting factors are above 0.97. Lower fitting factors are expected for the remaining systems where the neutron star component spin is higher than 0.05 or lower than -0.05, as the NSBH part of the bank does not have templates here. However the fit improves slightly at very high mass ratios (above about 70) since these systems are also high total mass systems which are recovered by the part of the bank where additional templates were placed with a higher (98\%) minimum match.}
\end{figure*}

In Fig.~\ref{fig:nsbhff}, we can see the fitting factors in the $M$-$\chi_\mathrm{eff}$ plane and as a function of mass ratio for NSBH aligned-spin SEOBNRv4\_ROM waveform approximants~\cite{SEOBNRv4ROM}. For 99.2\% of the simulations, the fitting factors are above 0.97. Lower fitting factors are expected for the remaining binary systems where the neutron star component spin is higher than 0.05 or lower than -0.05, as the NSBH part of the bank does not have templates here. However the fit improves slightly at very high mass ratios (above about 70) since these systems are also high total mass systems which are recovered by the part of the bank where additional templates were placed with a higher (98\%) minimum match.

\begin{figure*}[t]
\centering
\begin{minipage}[b]{0.47\textwidth}
\centering
  \includegraphics[width=\textwidth]{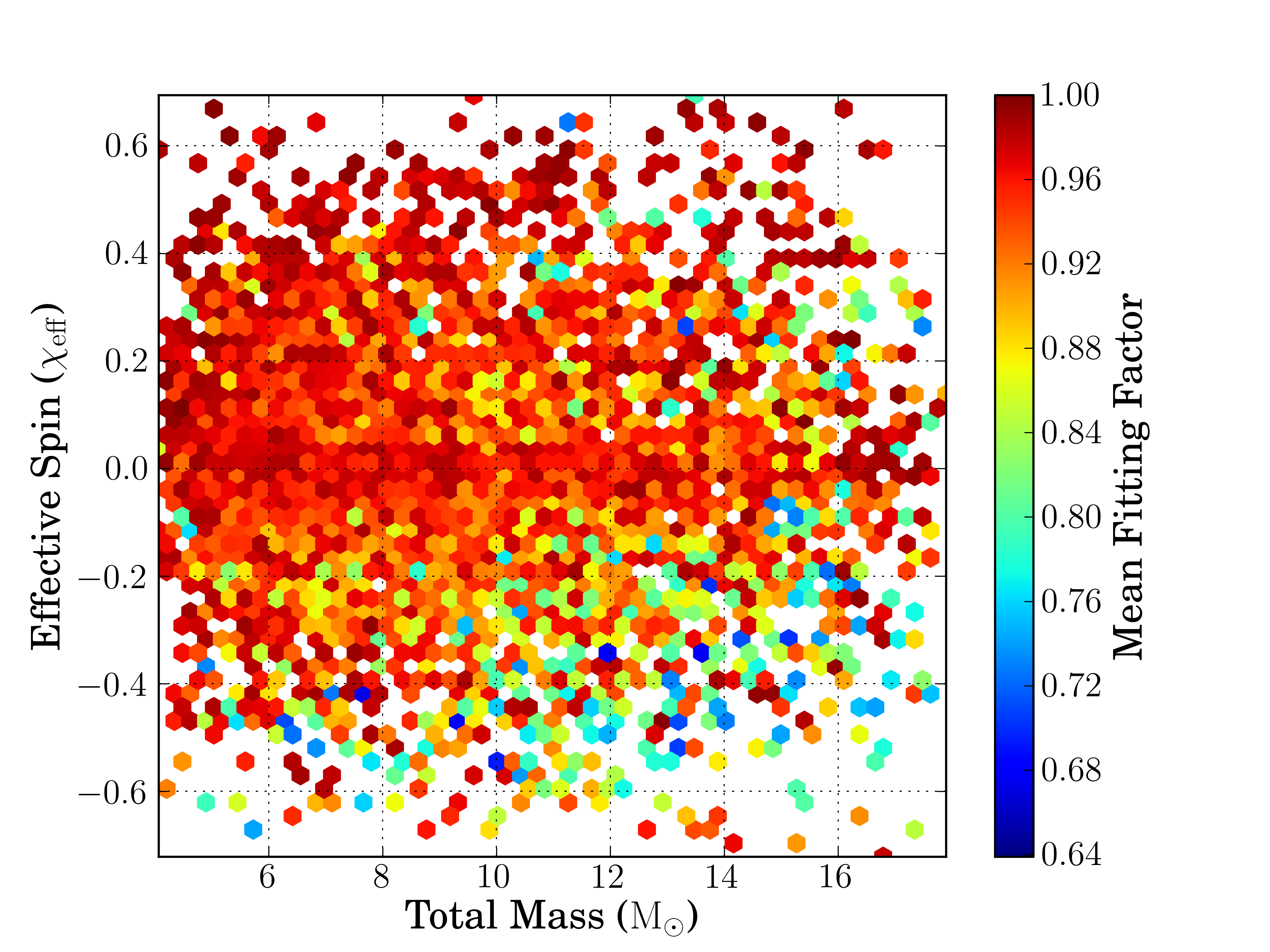}
  \end{minipage}
  \begin{minipage}[b]{0.47\textwidth}
  \centering
  \includegraphics[width=\textwidth]{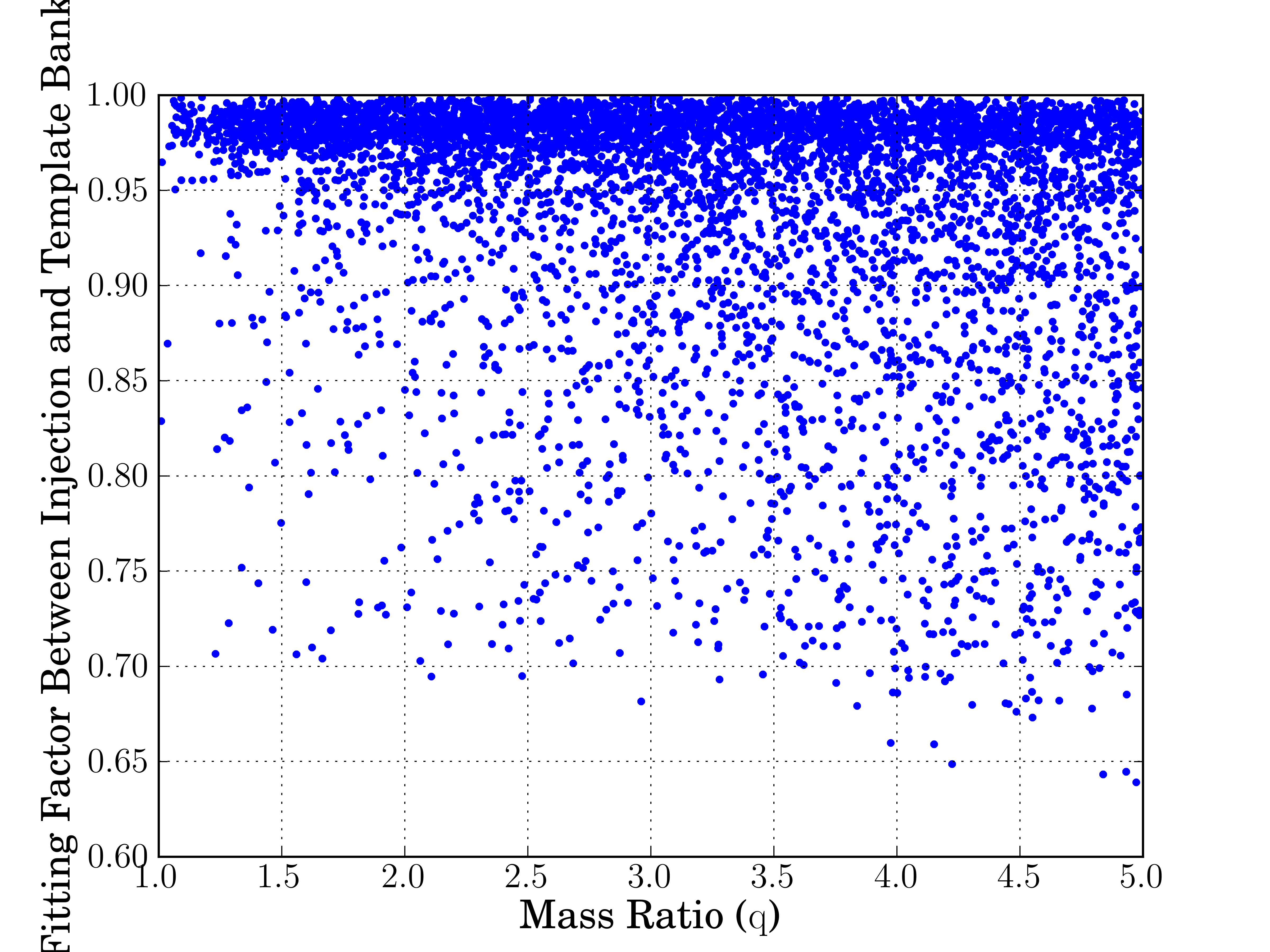}
  \end{minipage}
  \begin{minipage}[b]{0.47\textwidth}
  \centering
  \includegraphics[width=\textwidth]{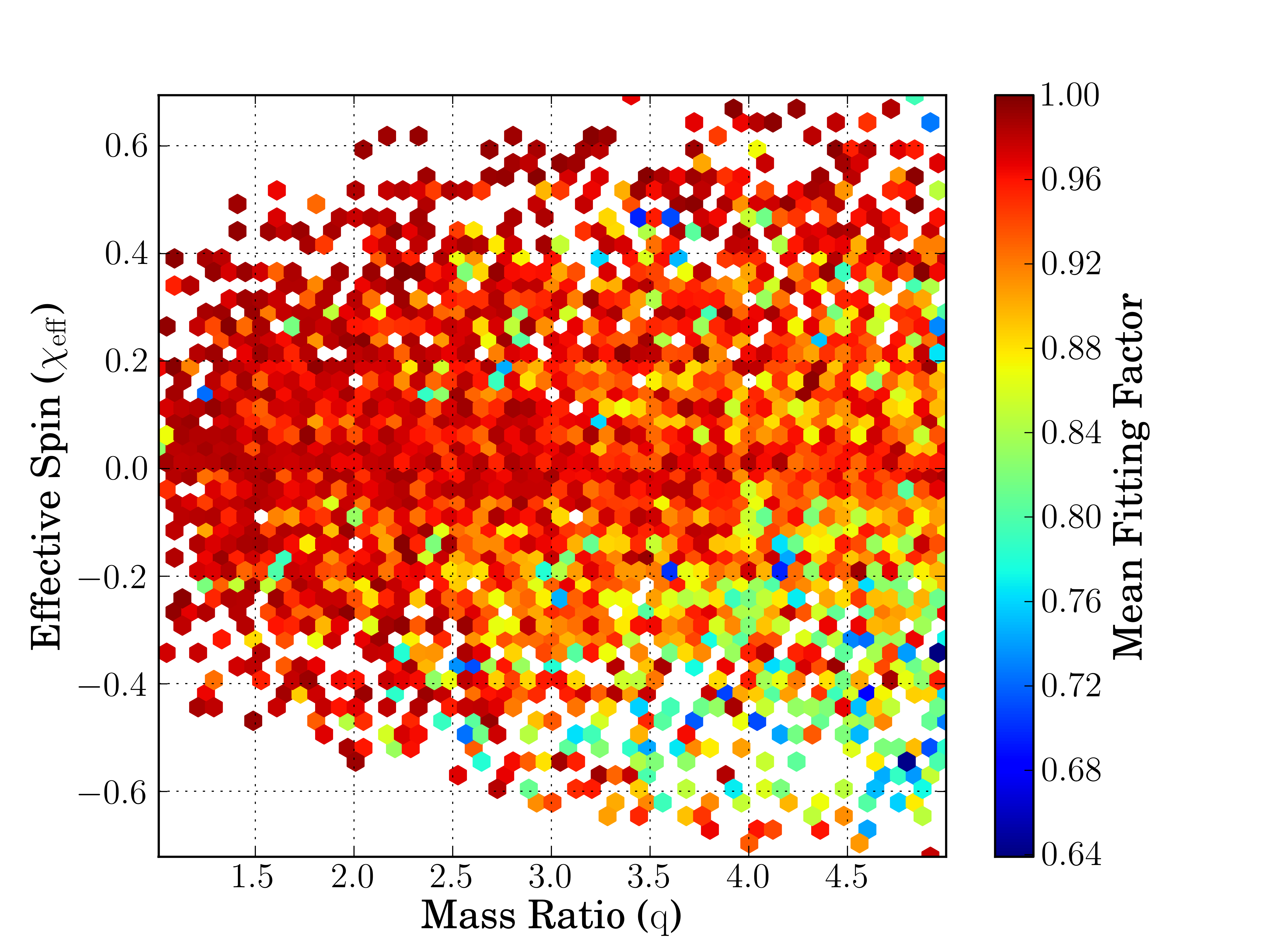}
  \end{minipage}
  \begin{minipage}[b]{0.47\textwidth}
  \centering
  \includegraphics[width=\textwidth]{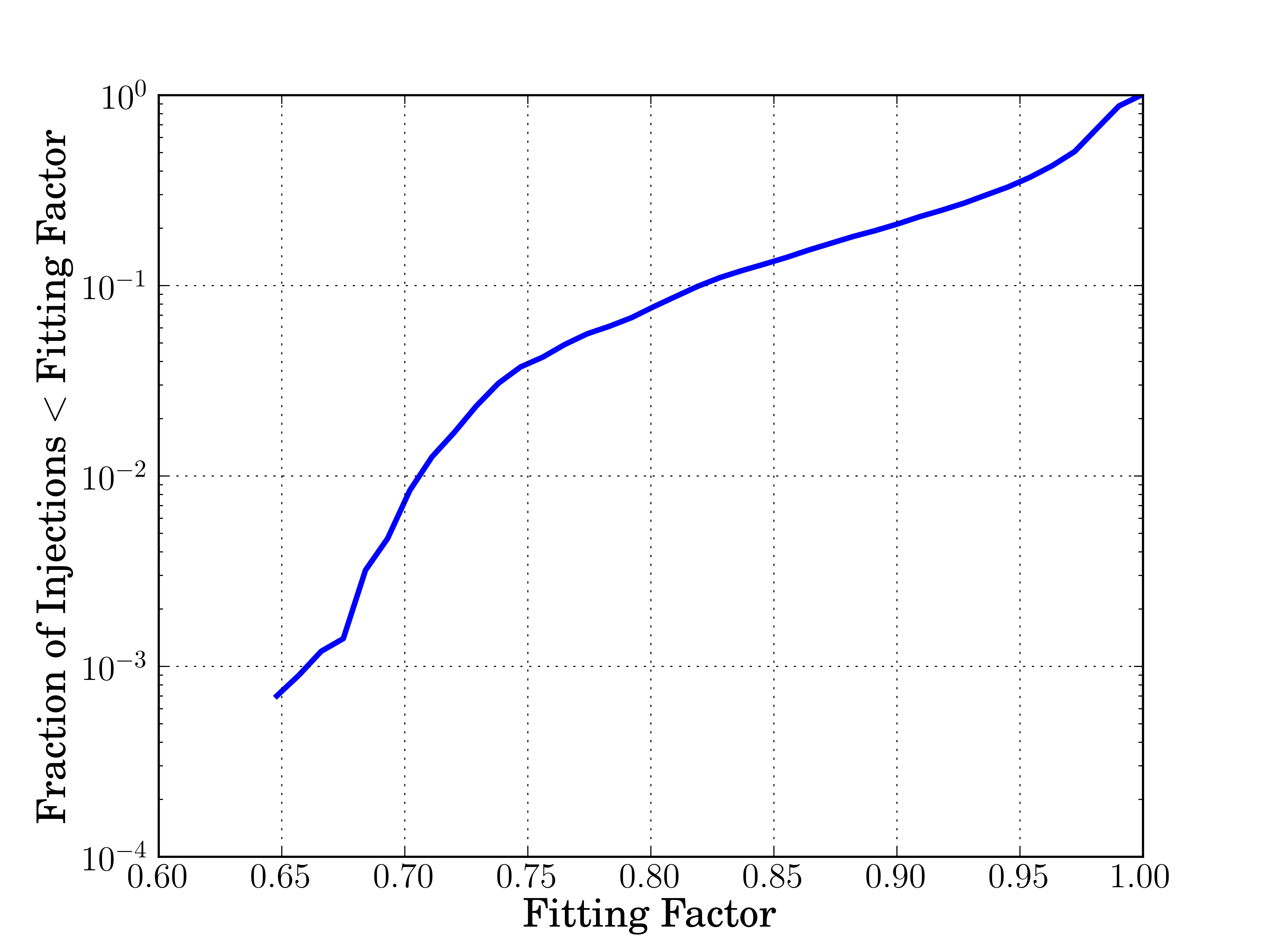}
  \end{minipage}
\caption{\label{fig:nsbhimrff}(Color online) Fitting factors in $M$-$\chi_\mathrm{eff}$ plane ({\it top left}), as a function of mass ratio (q) ({\it top right}) and in the $q$-$\chi_\mathrm{eff}$ plane ({\it bottom left}) for NSBH precessing IMRPhenomPv2 waveform approximants~\cite{IMRPhenomPv2}. We also include a cumulative histogram of the fitting factors ({\it bottom right}). Only 51.5\% of fitting factors are above 0.97 and fitting factors down to below 0.65 are present across this region. ({\it Top right}) We can see that the low fitting factors occur for systems with more extreme mass ratios, while ({\it top left}) tells us the fitting factors are lower for the highly spinning systems. This is what we expect, since our bank does not have templates that include precession effects, which depend on the mass ratio and the effective spin of the binary. The combined effect can be seen in the ({\it bottom left}) plot, where the increased mass ratio and $\chi_\mathrm{eff}$ shows lower fit.}
\end{figure*}

The precessing NSBH simulations using the IMRPhenomPv2 waveform~\cite{IMRPhenomPv2} in Fig.~\ref{fig:nsbhimrff} seem to be recovered with a lower fit for the highly precessing systems with high effective spins and mass ratios and only 51.5\% of the fitting factors are above 0.97. This is what we expect, since our bank does not have templates that include precession effects, which depend on the mass ratio and the effective spin of the binary. 

\begin{figure*}[t] 
\centering
\begin{minipage}[b]{0.47\textwidth}
\centering
  \includegraphics[width=\textwidth]{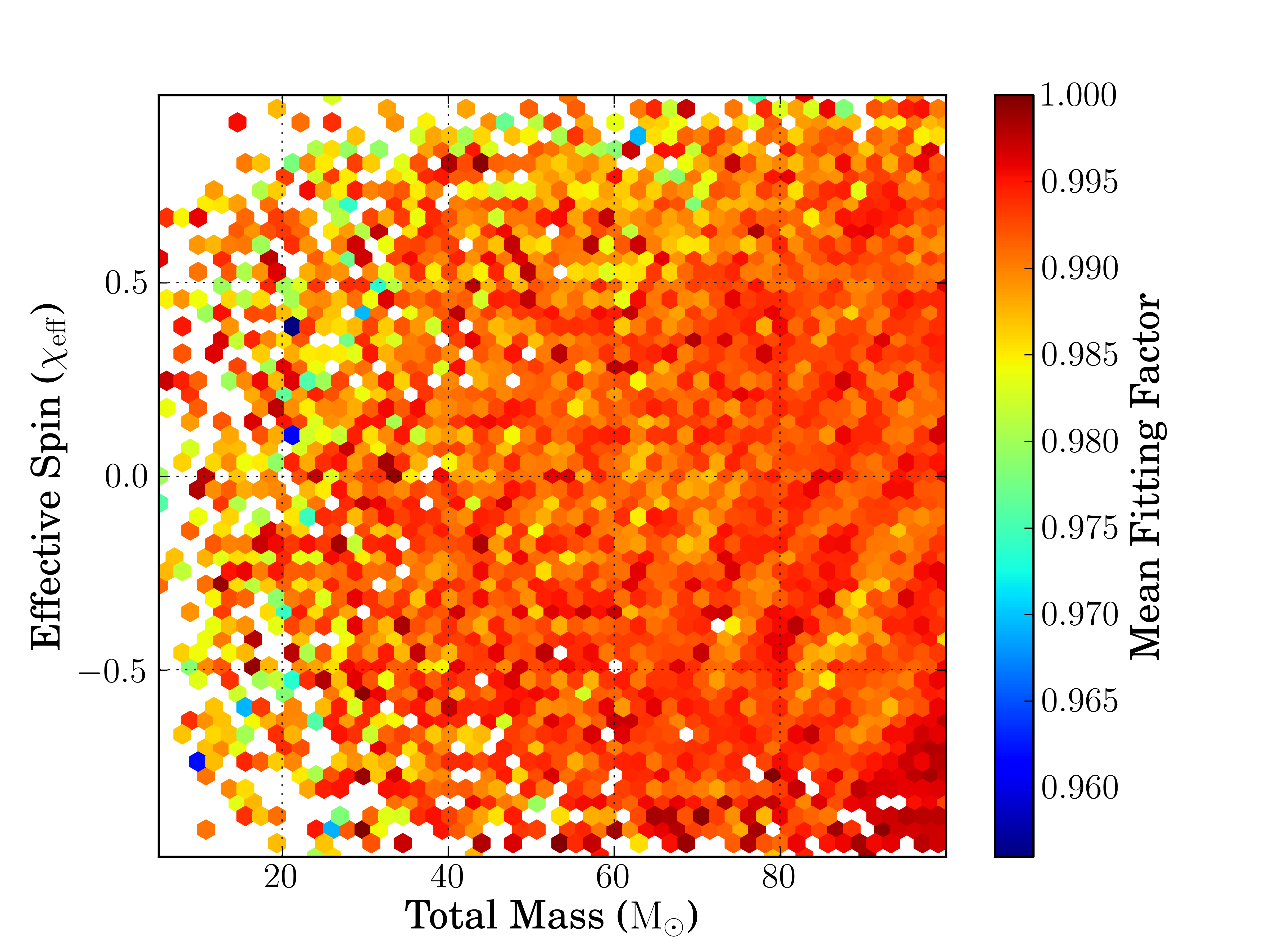}
  \end{minipage}
  \begin{minipage}[b]{0.47\textwidth}
  \centering
  \includegraphics[width=\textwidth]{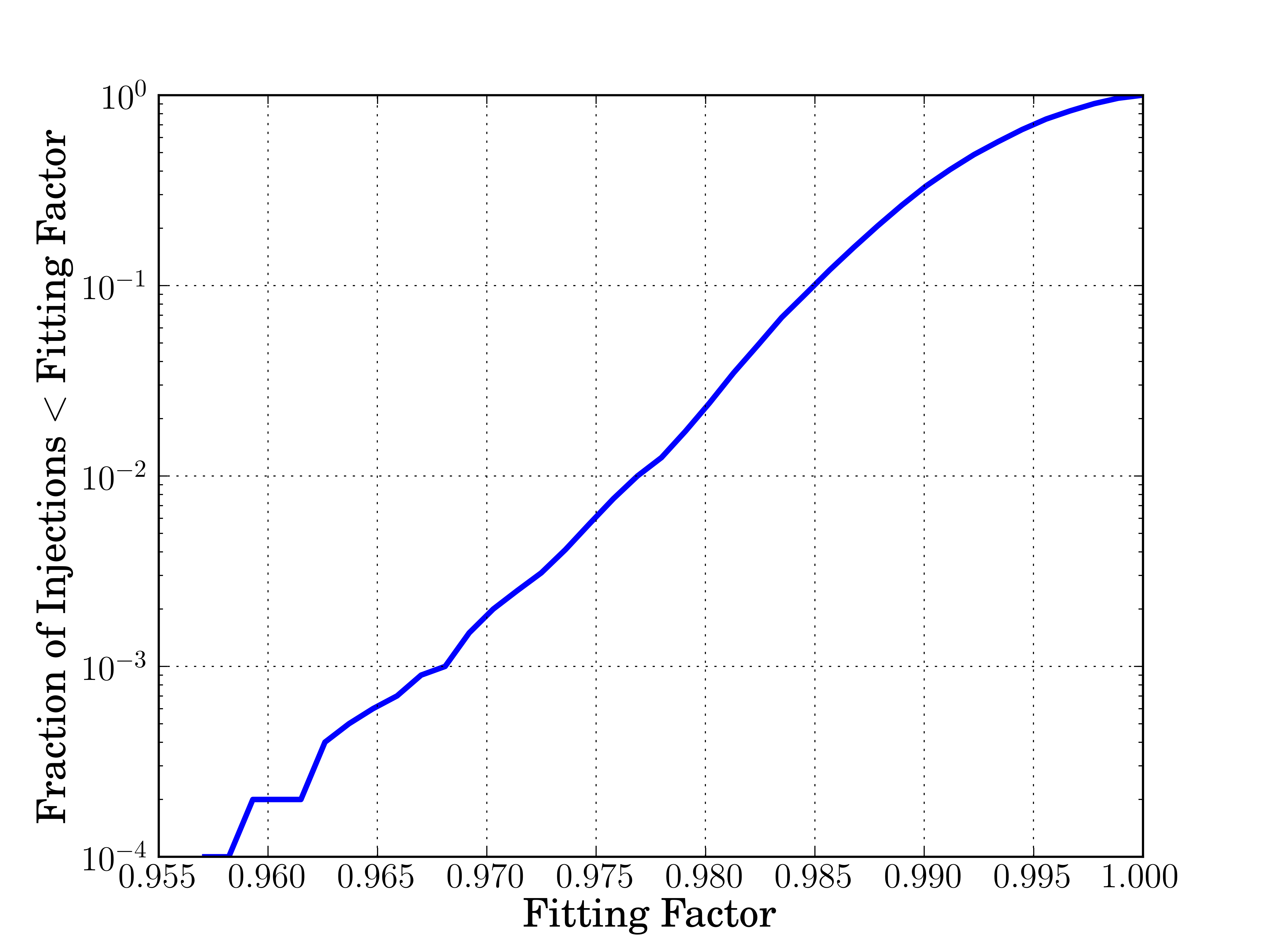}
  \end{minipage}
  \begin{minipage}[b]{0.47\textwidth}
  \centering
  \includegraphics[width=\textwidth]{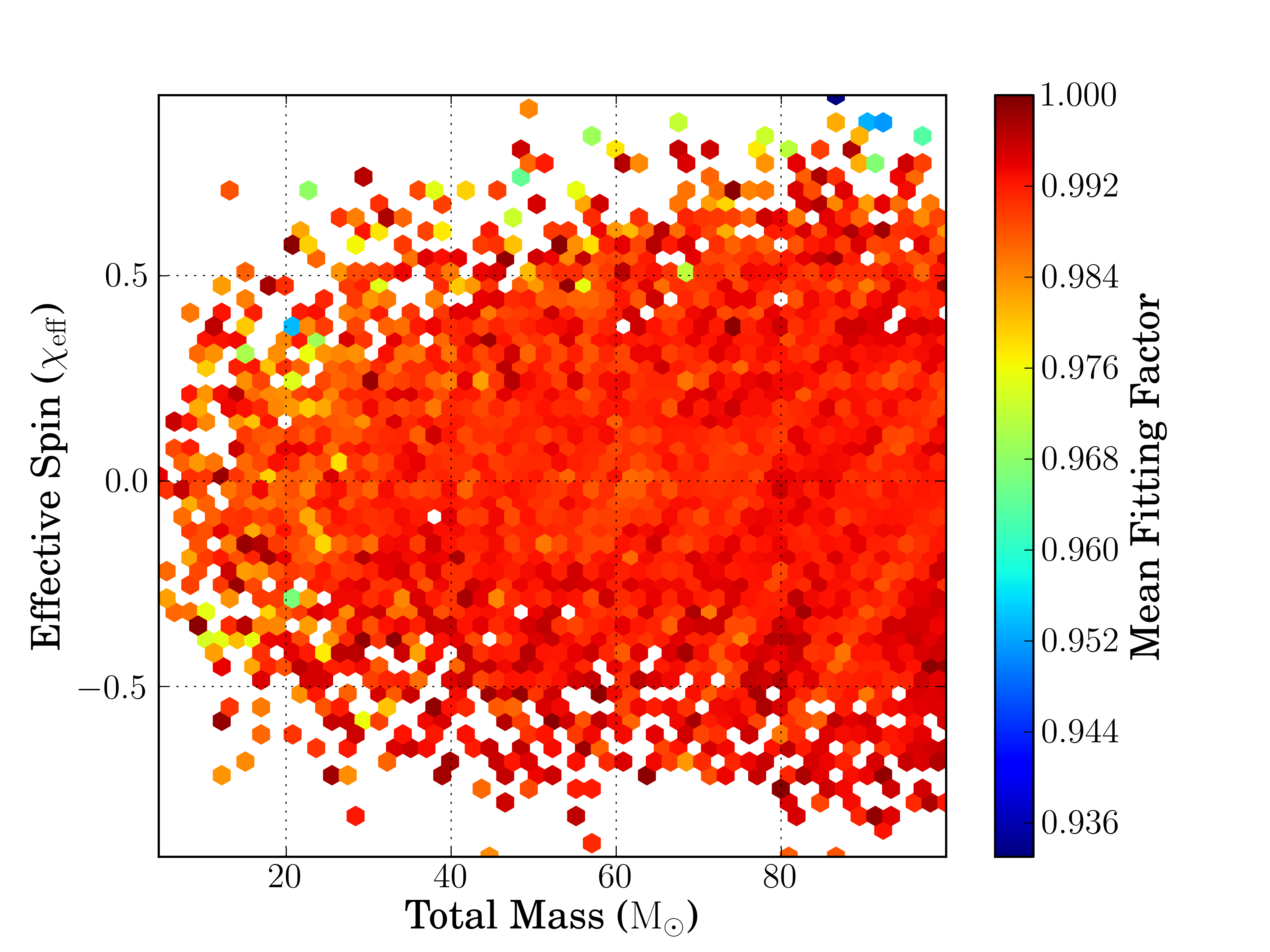}
  \end{minipage}
  \begin{minipage}[b]{0.47\textwidth}
  \centering
  \includegraphics[width=\textwidth]{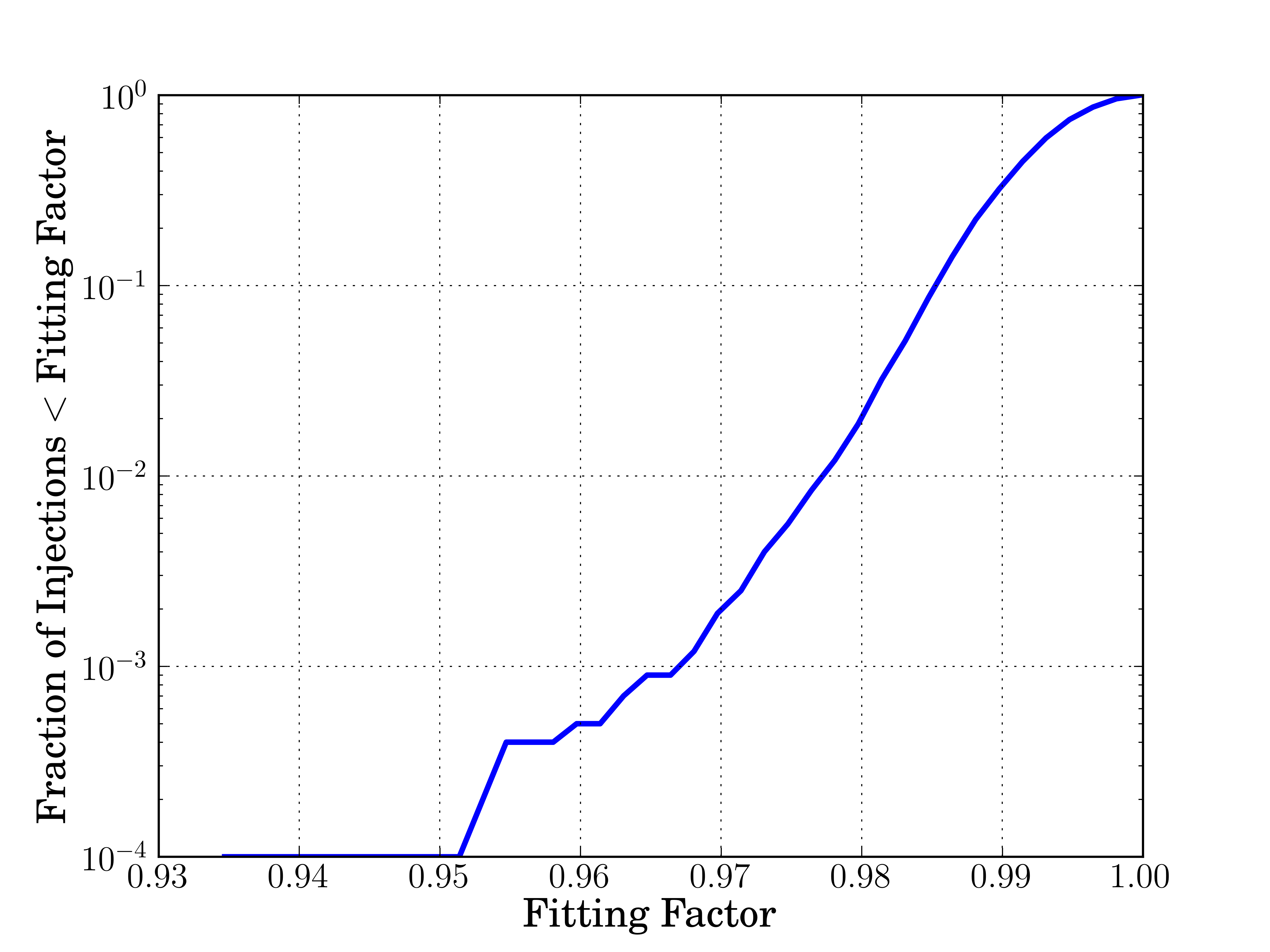}
  \end{minipage}
\caption{\label{fig:bbhff}(Color online) Fitting factors in $M$-$\chi_\mathrm{eff}$ plane for BBH aligned-spin SEOBNRv4\_ROM waveform approximants~\cite{SEOBNRv4ROM} ({\it top left}) and precessing-spin SEOBNRv2\_ROM\_DoubleSpin waveform approximants~\cite{SEOBNR_Double} ({\it bottom left}). We also include cumulative histograms of the fitting factors of the respective waveforms ({\it top and bottom right}). For the aligned-spin BBH systems, we recover 99.8\% and for the precessing BBH systems we recover 99.79\%  of the injected simulations fitting factors are above 0.97. }
\end{figure*}

\begin{figure*}[t]
\centering
\begin{minipage}[b]{0.47\textwidth}
\centering
  \includegraphics[width=\textwidth]{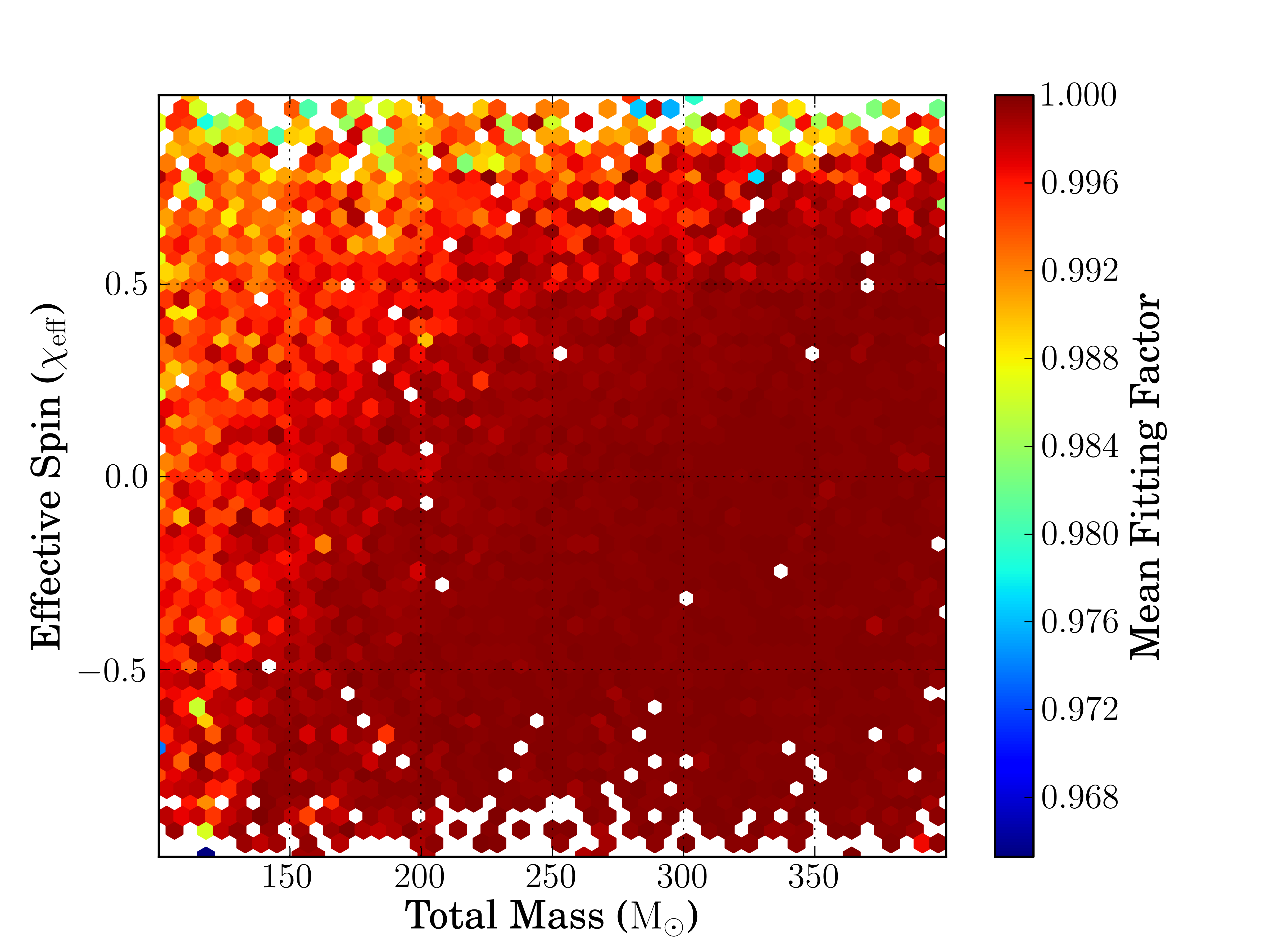}
  \end{minipage}
  \begin{minipage}[b]{0.47\textwidth}
  \centering
 \includegraphics[width=\textwidth]{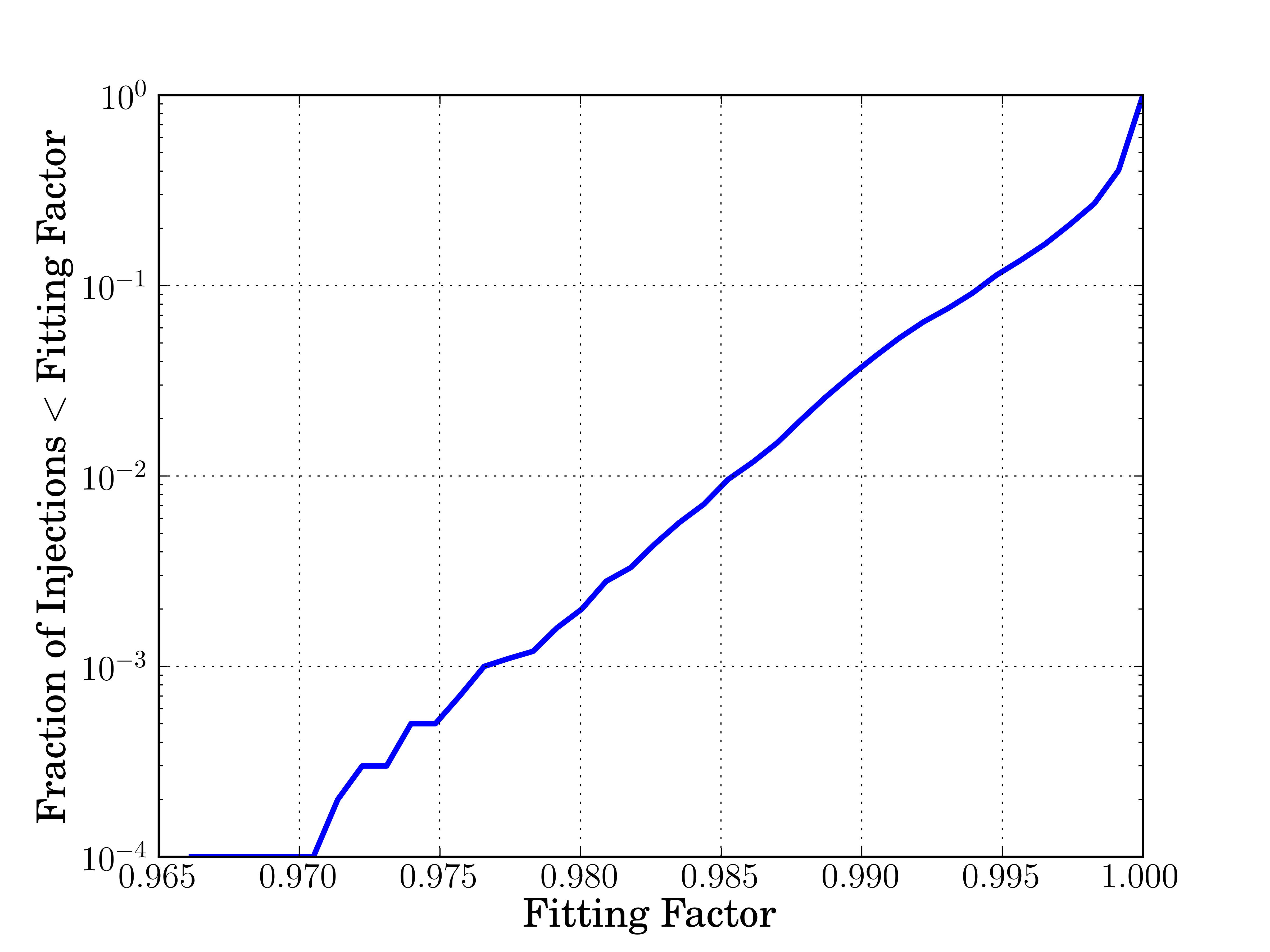}
  \end{minipage}
  \begin{minipage}[b]{0.47\textwidth}
  \centering
 \includegraphics[width=\textwidth]{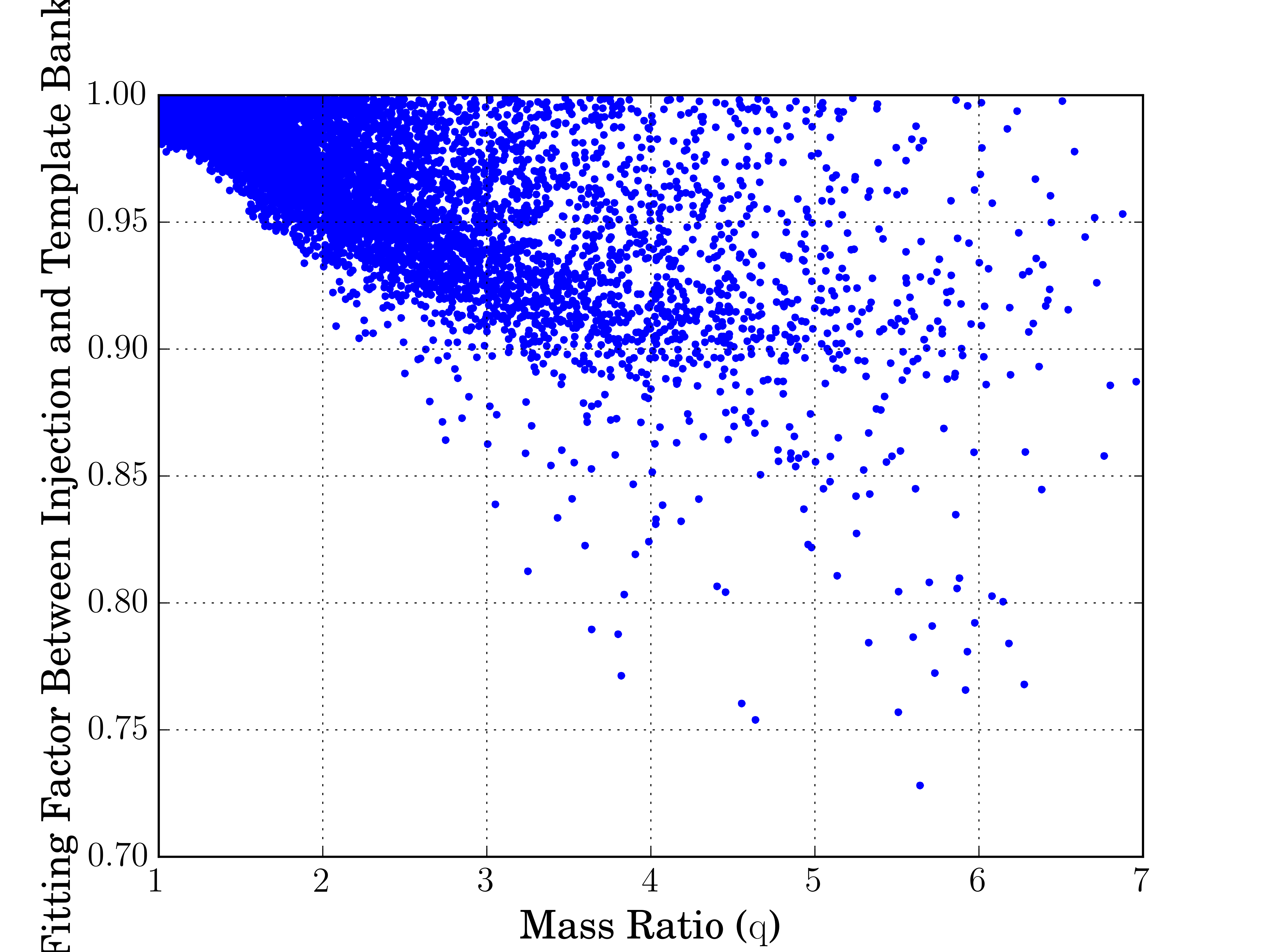}
  \end{minipage}
  \begin{minipage}[b]{0.47\textwidth}
  \centering
 \includegraphics[width=\textwidth]{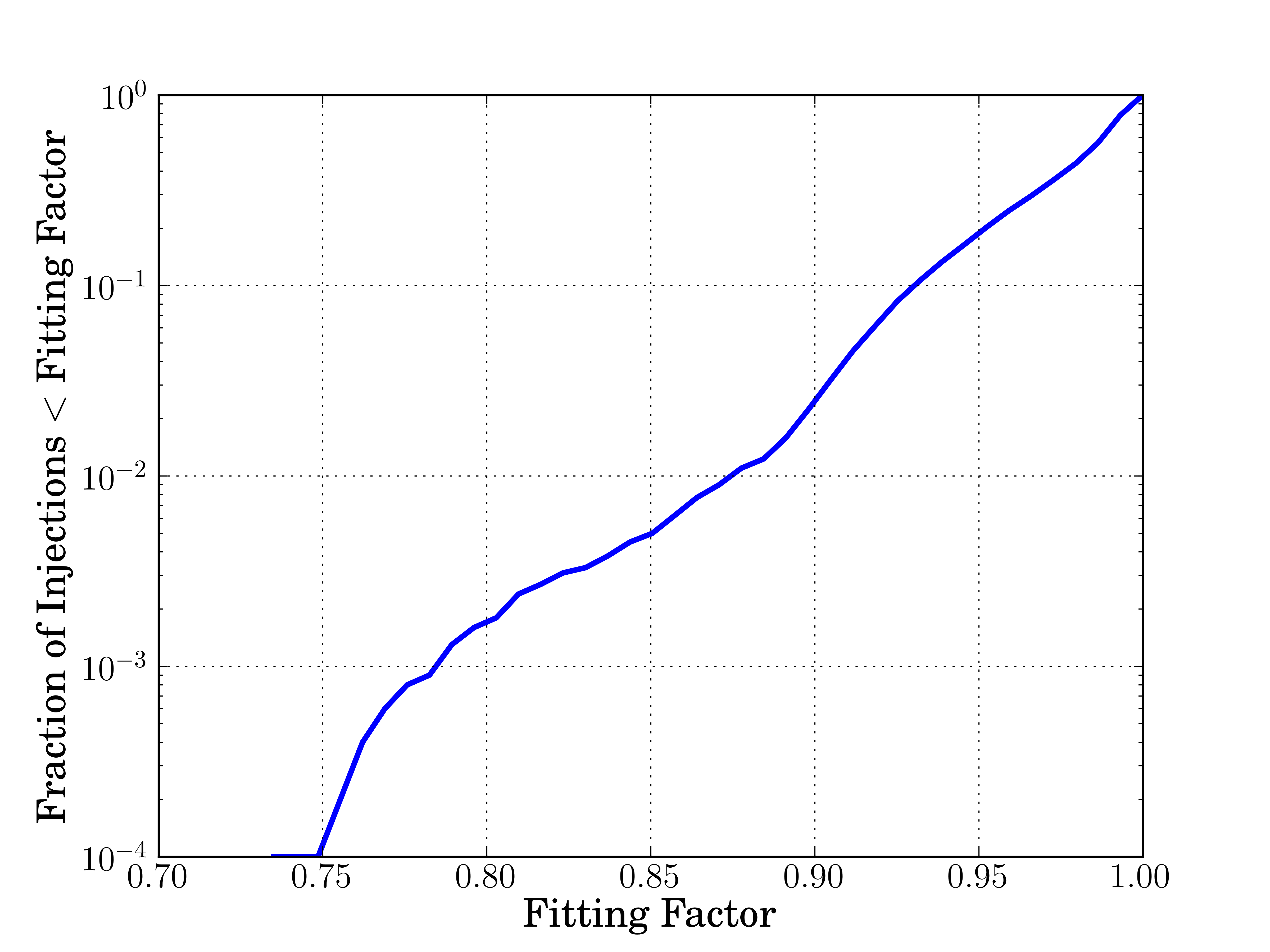}
  \end{minipage}
\caption{\label{fig:imbhff}(Color online) Fitting factors in $M$-$\chi_\mathrm{eff}$ plane for IMBHB aligned-spin SEOBNRv4\_ROM waveform approximants~\cite{SEOBNRv4ROM} ({\it top left}) and as a function of mass ratio for IMBHB non-spinning EOBNRv2HM waveform approximants~\cite{EOBNRv2HM} ({\it bottom left}). Cumulative histograms of the fitting factors of the respective waveforms is also included ({\it top and bottom right}). 99.99\% of fitting factors are above 0.97 for the recovery of aligned-spin SEOBNRv4\_ROM waveform approximants. Non-spinning EOBNRv2HM waveform approximants with higher-order modes can also be recovered by the search in the IMBHB region, despite template waveforms not including higher-order mode effects. We recover 66.6\% of these simulations with fitting factors above 0.97.}
\end{figure*}

In Figures~\ref{fig:bbhff} and~\ref{fig:imbhff}, we can see the fitting factors in $M$-$\chi_\mathrm{eff}$ plane for BBH and IMBHB aligned-spin SEOBNRv4\_ROM waveform approximants~\cite{SEOBNRv4ROM}, the precessing BBH SEOBNRv2\_ROM\_DoubleSpin waveform approximants~\cite{SEOBNR_Double} and as a function of mass ratio for IMBHB non-spinning EOBNRv2HM waveform approximants~\cite{EOBNRv2HM}. For the aligned-spin SEOBNRv4\_ROM waveform approximants, 99.8\% of the injected BBH simulations and 99.99\% of the IMBHB simulations are recovered with fitting factors above 0.97. The bank is constructed with SEOBNRv4\_ROM waveforms in the high mass region so fitting factors are expected to be at least as high as the required fitting factor of 0.97. Even though the bank does not include precessing templates, 99.79\% of the precessing-spin SEOBNRv2\_ROM\_DoubleSpin BBH waveform approximants are recovered with fitting factors above 0.97. Non-spinning EOBNRv2HM waveform approximants including modes higher than the fundamental 2,2 mode, can also be recovered by the search in the IMBHB region, despite template waveforms not including such higher-order mode effects. We recover 66.6\% of these simulations with fitting factors above 0.97 and the fit is seen to decline at higher mass ratios. This is because, the fitting factors have a dependency on mass ratios, as effects of higher-order modes become more significant at higher total mass and mass ratios~\cite{IMBHHM}. 

\section{Conclusion} \label{sec:conclusion}

We have presented the construction and effectualness of the aligned-spin online and offline template banks of
gravitational waveforms used by the GstLAL-based inspiral pipeline to analyze
data from the second observing run of Advanced LIGO and Virgo. The offline bank expands
upon the parameter space covered during the first observing run, including coverage
for merging compact binary systems with total mass between
2\,$\msun$ and 400\,$\msun$ and mass ratios between 1 and 97.988, thus expanding into the intermediate-mass
binary black hole range. The bank requires that templates with component masses less than 2\,$\msun$ have (anti-)aligned spins between $\pm0.05$ while component masses greater than 2\,$\msun$ have allowed (anti-)aligned between
$\pm0.999$. Despite this aligned-spin constraint, we find that the bank can recover some precessing-spin systems.
Additionally, higher-order mode effects are not included in the template waveforms but the bank can recover some non-spinning IMBHB waveforms with higher-order modes.

We expect to be sensitive to BNSs with component masses 1-3 $\msun$ and (anti-) aligned spins between $\pm0.05$. Our sensitivity to precessing BNS systems with spins (anti-) aligned above 0.05 or below -0.05 is limited as discussed in Sect.~\ref{sec:effectual} and in Sect.~\ref{subsec:astro}. We are sensitive to NSBH systems, with component masses in the range 1-3 and 3-97 $\msun$ and (anti-)aligned spins between $\pm0.05$ for the neutron star component and between $\pm0.99$ for the black hole component. Precessing NSBH systems and ones with component spins outside this range are recovered with poorer fits. We are most sensitive to BBH systems with component masses in the range 2-99  $\msun$ and (anti-)aligned spins in the range $\pm0.99$ and have poorer recovery for precessing systems. We are the most sensitive to IMBHB systems with component masses between 1-399 $\msun$, total mass in the range 100-400 $\msun$ and (anti-)aligned spins in the range $\pm0.998$. Recovery for IMBHB waveforms including effects of higher order modes, is seen to fall with increase in mass ratios. Higher-order modes have higher frequency content and will be within the sensitive frequency band of LIGO and Virgo for IMBHB signals. Hence it will become important to include templates containing higher-order mode contributions, in order to increase the sensitivity of the search towards heavier mass systems~\cite{IMBHHM}.

The online and offline banks played key roles in the discoveries of O2~\cite{GW170104, GW170608, GW170814, GW170817} and the offline bank was used in the deep GstLAL reanalysis of O1 and O2~\cite{O1-O2Catalog}. The present bank however contains templates that assume the spins of the stars to be aligned with their orbital angular momentum. They do not include precession. Hence we see precessing waveforms are recovered with a lower fit with the templates that recover them. The bank also does not include templates with higher order modes in their waveforms which has been seen to reduce our sensitivity to them. The bank includes templates for systems with mass ratios going from 1-97.988 and it could be expanded to include contributions from higher order modes. From the construction of this template bank, we also learned that we need optimal coverage in the region representing the higher mass IMBHB systems to better estimate the background statistics. We additionally presented a new method that combines a stochastic method with a grid-bank method to better isolate noisy templates at the high mass region of the bank. This allowed for better grouping of templates when performing background estimation. A more careful layout and grouping of templates has been implemented in our template bank for O3, which will be covered in a future work. The experience gained in designing these banks has informed the construction of the template bank which is being used for the third observing run of Advanced LIGO and Virgo.
	
\section{Acknowledgements} \label{sec:ack}

We thank the LIGO-Virgo Scientific Collaboration for access to data. LIGO was
constructed by the California Institute of Technology and Massachusetts
Institute of Technology with funding from the National Science Foundation (NSF)
and operates under cooperative agreement PHY-0757058. We thank Satya Mohapatra for the helpful comments and suggestions. We also thank Graham Woan for helping with the review of the template bank used for O2. We gratefully acknowledge the support by NSF grant PHY-1607585 for JC, PB, DC and DM, who was also partly supported by PHY 14-54389, ACI 16-42391, OAC 18-41480. SC is supported by the research programme of the Netherlands Organisation for Scientific Research (NWO). SS was supported in part by the Eberly Research Funds of Penn State, The Pennsylvania State University, University Park, PA 16802, USA. HF was supported by the Natural Sciences and Engineering Research Council of Canada (NSERC). CH was supported in part by the NSF through PHY-1454389.  Funding for this project was provided by the Charles E.  Kaufman Foundation of The Pittsburgh Foundation. TGFL was partially supported by a grant from the Research Grants Council of the Hong Kong (Project No. CUHK 14310816 and CUHK 24304317) and the Direct Grant for Research from the Research Committee of the Chinese University of Hong Kong. MW was supported by NSF grant PHY-1607178. We thank the computational resources provided by Leonard E Parker Center for Gravitation, Cosmology and Astrophysics at the University of Wisconsin-Milwaukee and supported by National Science Foundation Grants PHY-1626190 and PHY-1700765. We are also grateful for the computational resources provided by the Pennsylvania State University's Institute for CyberScience Advanced Cyber-Infrastructure (ICS-ACI). The authors are also grateful for computational resources provided by the LIGO Laboratory and supported by National Science Foundation Grants PHY-0757058 and PHY-0823459. This paper carries LIGO Document Number LIGO-P1700412. 


\begin{thebibliography}{49}%
\makeatletter
\providecommand \@ifxundefined [1]{%
 \@ifx{#1\undefined}
}%
\providecommand \@ifnum [1]{%
 \ifnum #1\expandafter \@firstoftwo
 \else \expandafter \@secondoftwo
 \fi
}%
\providecommand \@ifx [1]{%
 \ifx #1\expandafter \@firstoftwo
 \else \expandafter \@secondoftwo
 \fi
}%
\providecommand \natexlab [1]{#1}%
\providecommand \enquote  [1]{``#1''}%
\providecommand \bibnamefont  [1]{#1}%
\providecommand \bibfnamefont [1]{#1}%
\providecommand \citenamefont [1]{#1}%
\providecommand \href@noop [0]{\@secondoftwo}%
\providecommand \href [0]{\begingroup \@sanitize@url \@href}%
\providecommand \@href[1]{\@@startlink{#1}\@@href}%
\providecommand \@@href[1]{\endgroup#1\@@endlink}%
\providecommand \@sanitize@url [0]{\catcode `\\12\catcode `\$12\catcode
  `\&12\catcode `\#12\catcode `\^12\catcode `\_12\catcode `\%12\relax}%
\providecommand \@@startlink[1]{}%
\providecommand \@@endlink[0]{}%
\providecommand \url  [0]{\begingroup\@sanitize@url \@url }%
\providecommand \@url [1]{\endgroup\@href {#1}{\urlprefix }}%
\providecommand \urlprefix  [0]{URL }%
\providecommand \Eprint [0]{\href }%
\providecommand \doibase [0]{http://dx.doi.org/}%
\providecommand \selectlanguage [0]{\@gobble}%
\providecommand \bibinfo  [0]{\@secondoftwo}%
\providecommand \bibfield  [0]{\@secondoftwo}%
\providecommand \translation [1]{[#1]}%
\providecommand \BibitemOpen [0]{}%
\providecommand \bibitemStop [0]{}%
\providecommand \bibitemNoStop [0]{.\EOS\space}%
\providecommand \EOS [0]{\spacefactor3000\relax}%
\providecommand \BibitemShut  [1]{\csname bibitem#1\endcsname}%
\let\auto@bib@innerbib\@empty
\bibitem [{\citenamefont {Harry}\ \emph {et~al.}(2010)\citenamefont {Harry}
  \emph {et~al.}}]{AdvLIGO1}%
  \BibitemOpen
  \bibfield  {author} {\bibinfo {author} {\bibfnamefont {G.~M}~\bibnamefont
  {Harry}} \emph {et~al.} (\bibinfo {collaboration} {The LIGO Scientific Collaboration}),\ }\href {\doibase 10.1088/0264-9381/27/8/084006} {\bibfield
  {journal} {\bibinfo  {journal} {Classical and Quantum Gravity}\ }\textbf {\bibinfo {volume}
  {27}},\ \bibinfo {pages} {084006} (\bibinfo {year} {2010})}\BibitemShut
  {NoStop}%
\bibitem [{\citenamefont {Aasi}\ \emph
  {et~al.}(2015{\natexlab{a}})\citenamefont {Aasi} \emph
  {et~al.}}]{AdvLIGO2}%
  \BibitemOpen
  \bibfield  {author} {\bibinfo {author} {\bibfnamefont {J}\ \bibnamefont {Aasi}} \emph {et~al.} (\bibinfo {collaboration} {The LIGO Scientific Collaboration}),\ }\href {\doibase 10.1088/0264-9381/32/7/074001} {\bibfield  {journal} {\bibinfo  {journal} {Classical and Quantum Gravity}\ }\textbf {\bibinfo {volume} {32}},\ \bibinfo {pages} {074001} (\bibinfo {year} {2015})}\BibitemShut {NoStop}%
\bibitem [{\citenamefont {Acernese}\ \emph {et~al.}(2014)\citenamefont
  {Acernese}, \citenamefont {Agathos}, \citenamefont {Agatsuma} \emph
  {et~al.}}]{AdvVirgo}%
  \BibitemOpen
  \bibfield  {author} {\bibinfo {author} {\bibfnamefont {F.}~\bibnamefont
  {Acernese}}, \bibinfo {author} {\bibfnamefont {M.}~\bibnamefont {Agathos}},
  \bibinfo {author} {\bibfnamefont {K.}~\bibnamefont {Agatsuma}},  \emph
  {et~al.},\ }\href {\doibase 10.1088/0264-9381/32/2/024001} {\bibfield  {journal}
  {\bibinfo  {journal} {Classical and Quantum Gravity}\ }\textbf {\bibinfo {volume} {32}},\
  \bibinfo {pages} {024001} (\bibinfo {year} {2014})}\BibitemShut {NoStop}%
\bibitem [{\citenamefont {Abbott}\ \emph
  {et~al.}(2016{\natexlab{a}})\citenamefont {Abbott} \emph
  {et~al.}}]{GW150914}%
  \BibitemOpen
  \bibfield  {author} {\bibinfo {author} {\bibfnamefont {B.~P.}\ \bibnamefont
  {Abbott}} \emph {et~al.} (\bibinfo {collaboration} {LIGO Scientific
  Collaboration and Virgo Collaboration}),\ }\href {\doibase
  10.1103/PhysRevLett.116.061102} {\bibfield  {journal} {\bibinfo  {journal}
  {Phys. Rev. Lett.}\ }\textbf {\bibinfo {volume} {116}},\ \bibinfo {pages}
  {061102} (\bibinfo {year} {2016}{\natexlab{a}})}\BibitemShut {NoStop}%
\bibitem [{\citenamefont {Albert}\ \emph {et~al.}(2017)\citenamefont {Albert}
  \emph {et~al.}}]{GW151226}%
  \BibitemOpen
  \bibfield  {author} {\bibinfo {author} {\bibfnamefont {A.}~\bibnamefont
  {Albert}} \emph {et~al.} (\bibinfo {collaboration} {ANTARES Collaboration and
  IceCube Collaboration and LIGO Scientific Collaboration, and Virgo
  Collaboration}),\ }\href {\doibase 10.1103/PhysRevD.96.022005} {\bibfield
  {journal} {\bibinfo  {journal} {Phys. Rev. D}\ }\textbf {\bibinfo {volume}
  {96}},\ \bibinfo {pages} {022005} (\bibinfo {year} {2017})}\BibitemShut
  {NoStop}%
\bibitem [{\citenamefont {Abbott}\ \emph
  {et~al.}(2017{\natexlab{a}})\citenamefont {Abbott} \emph
  {et~al.}}]{GW170104}%
  \BibitemOpen
  \bibfield  {author} {\bibinfo {author} {\bibfnamefont {B.~P.}\ \bibnamefont
  {Abbott}} \emph {et~al.} (\bibinfo {collaboration} {LIGO Scientific and Virgo
  Collaboration}),\ }\href {\doibase 10.1103/PhysRevLett.118.221101} {\bibfield
   {journal} {\bibinfo  {journal} {Phys. Rev. Lett.}\ }\textbf {\bibinfo
  {volume} {118}},\ \bibinfo {pages} {221101} (\bibinfo {year}
  {2017}{\natexlab{a}})}\BibitemShut {NoStop}%
\bibitem [{\citenamefont {Abbott}\ \emph
  {et~al.}(2017{\natexlab{b}})\citenamefont {Abbott} \emph
  {et~al.}}]{GW170608}%
  \BibitemOpen
  \bibfield  {author} {\bibinfo {author} {\bibfnamefont {B.~P.}\ \bibnamefont
  {Abbott}} \emph {et~al.},\ }\href
  {http://stacks.iop.org/2041-8205/851/i=2/a=L35} {\bibfield  {journal}
  {\bibinfo  {journal} {The Astrophysical Journal Letters}\ }\textbf {\bibinfo
  {volume} {851}},\ \bibinfo {pages} {L35} (\bibinfo {year}
  {2017}{\natexlab{b}})}\BibitemShut {NoStop}%
\bibitem [{\citenamefont {Abbott}\ \emph
  {et~al.}(2017{\natexlab{c}})\citenamefont {Abbott} \emph
  {et~al.}}]{GW170814}%
  \BibitemOpen
  \bibfield  {author} {\bibinfo {author} {\bibfnamefont {B.~P.}\ \bibnamefont
  {Abbott}} \emph {et~al.} (\bibinfo {collaboration} {LIGO Scientific
  Collaboration and Virgo Collaboration}),\ }\href {\doibase
  10.1103/PhysRevLett.119.141101} {\bibfield  {journal} {\bibinfo  {journal}
  {Phys. Rev. Lett.}\ }\textbf {\bibinfo {volume} {119}},\ \bibinfo {pages}
  {141101} (\bibinfo {year} {2017}{\natexlab{c}})}\BibitemShut {NoStop}%
\bibitem [{\citenamefont {Abbott}\ \emph
  {et~al.}(2017{\natexlab{d}})\citenamefont {Abbott} \emph
  {et~al.}}]{GW170817}%
  \BibitemOpen
  \bibfield  {author} {\bibinfo {author} {\bibfnamefont {B.~P.}\ \bibnamefont
  {Abbott}} \emph {et~al.} (\bibinfo {collaboration} {LIGO Scientific
  Collaboration and Virgo Collaboration}),\ }\href {\doibase
  10.1103/PhysRevLett.119.161101} {\bibfield  {journal} {\bibinfo  {journal}
  {Phys. Rev. Lett.}\ }\textbf {\bibinfo {volume} {119}},\ \bibinfo {pages}
  {161101} (\bibinfo {year} {2017}{\natexlab{d}})}\BibitemShut {NoStop}%
\bibitem [{\citenamefont {Messick}\ \emph {et~al.}(2017)\citenamefont
  {Messick}, \citenamefont {Blackburn}, \citenamefont {Brady} \emph
  {et~al.}}]{O1Methods}%
  \BibitemOpen
  \bibfield  {author} {\bibinfo {author} {\bibfnamefont {C.}~\bibnamefont
  {Messick}}, \bibinfo {author} {\bibfnamefont {K.}~\bibnamefont {Blackburn}},
  \bibinfo {author} {\bibfnamefont {P.}~\bibnamefont {Brady}},  \emph
  {et~al.},\ }\href {\doibase 10.1103/PhysRevD.95.042001} {\bibfield  {journal}
  {\bibinfo  {journal} {Phys. Rev. D}\ }\textbf {\bibinfo {volume} {95}},\
  \bibinfo {pages} {042001} (\bibinfo {year} {2017})}\BibitemShut {NoStop}%
\bibitem [{\citenamefont {Klimenko}\ \emph {et~al.}(2008)\citenamefont
  {Klimenko}, \citenamefont {Yakushin}, \citenamefont {Mercer} \emph
  {et~al.}}]{cWB}%
  \BibitemOpen
  \bibfield  {author} {\bibinfo {author} {\bibfnamefont {S.}~\bibnamefont
  {Klimenko}}, \bibinfo {author} {\bibfnamefont {I.}~\bibnamefont {Yakushin}},
  \bibinfo {author} {\bibfnamefont {A.}~\bibnamefont {Mercer}},  \emph
  {et~al.},\ }\href {\doibase 10.1088/0264-9381/25/11/114029} {\bibfield  {journal}
  {\bibinfo  {journal} {Classical and Quantum Gravity}\ }\textbf {\bibinfo {volume} {25}},\
  \bibinfo {pages} {114029} (\bibinfo {year} {2008})}\BibitemShut {NoStop}%
\bibitem [{\citenamefont {Kumar}\ \emph {et~al.}(2014)\citenamefont
  {Kumar}, \citenamefont {MacDonald}, \citenamefont {Brown} \emph
  {et~al.}}]{nrbank}%
  \BibitemOpen
  \bibfield  {author} {\bibinfo {author} {\bibfnamefont {P.}~\bibnamefont
  {Kumar}}, \bibinfo {author} {\bibfnamefont {I.}~\bibnamefont {MacDonald}},
  \bibinfo {author} {\bibfnamefont {D.}~\bibnamefont {Brown}},  \emph
  {et~al.},\ }\href {\doibase 10.1103/PhysRevD.89.042002} {\bibfield  {journal}
  {\bibinfo  {journal} {Phys. Rev. D}\ }\textbf {\bibinfo {volume} {89}},\
  \bibinfo {pages} {042002} (\bibinfo {year} {2014})}\BibitemShut {NoStop}%
\bibitem [{\citenamefont {Lynch}\ \emph {et~al.}(2017)\citenamefont
  {Lynch}, \citenamefont {Vitale}, \citenamefont {Essick} \emph
  {et~al.}}]{LALinf}%
  \BibitemOpen
  \bibfield  {author} {\bibinfo {author} {\bibfnamefont {R.}~\bibnamefont
  {Lynch}}, \bibinfo {author} {\bibfnamefont {S.}~\bibnamefont {Vitale}},
  \bibinfo {author} {\bibfnamefont {R.}~\bibnamefont {Essick}},  \emph
  {et~al.},\ }\href {\doibase 10.1103/PhysRevD.95.104046} {\bibfield  {journal}
  {\bibinfo  {journal} {Phys. Rev. D}\ }\textbf {\bibinfo {volume} {95}},\
  \bibinfo {pages} {104046} (\bibinfo {year} {2017})}\BibitemShut {NoStop}%
\bibitem [{\citenamefont {Cornish}\ \emph {et~al.}(2015)\citenamefont
  {Cornish} \emph
  {et~al.}}]{Bayeswave}%
  \BibitemOpen
  \bibfield  {author} {\bibinfo {author} {\bibfnamefont {N.}~\bibnamefont
  {Cornish}},  \emph
  {et~al.},\ }\href {\doibase 10.1088/0264-9381/32/13/135012} {\bibfield  {journal}
  {\bibinfo  {journal} {Classical and Quantum Gravity}\ }\textbf {\bibinfo {volume} {32}},\
  \bibinfo {pages} {135012} (\bibinfo {year} {2015})}\BibitemShut {NoStop}%
\bibitem [{\citenamefont {Abbott, B. P}\ \emph {et~al.}(2019)\citenamefont
  {Abbott, B. P}, \citenamefont { Abbott, R}, \citenamefont {Abbott, T. D} \emph
  {et~al.}}]{IMBHO1-O2}%
  \BibitemOpen
  \bibfield  {author} {\bibinfo {author} {\bibfnamefont {B.P}~\bibnamefont
  {Abbott}}, \bibinfo {author} {\bibfnamefont {R.}~\bibnamefont {Abbott}},
  \bibinfo {author} {\bibfnamefont {T.D}~\bibnamefont {Abbott}},  \emph
  {et~al.},\ }\href {\doibase 10.1103/PhysRevD.100.064064} {\bibfield  {journal}
  {\bibinfo  {journal} {Phys. Rev. D}\ }\textbf {\bibinfo {volume} {100}},\
  \bibinfo {pages} {064064} (\bibinfo {year} {2019})}\BibitemShut {NoStop}%
\bibitem [{\citenamefont {Cannon}\ \emph {et~al.}()\citenamefont {Cannon},
  \citenamefont {Hanna} \emph {et~al.}}]{gstlal}%
  \BibitemOpen
  \bibfield  {author} {\bibinfo {author} {\bibfnamefont {K.}~\bibnamefont
  {Cannon}}, \bibinfo {author} {\bibfnamefont {C.}~\bibnamefont {Hanna}},
  \emph {et~al.},\ }\href@noop {} {\enquote {\bibinfo {title} {Gstlal},}\
  }\bibinfo {howpublished}
  {\url{https://www.lsc-group.phys.uwm.edu/daswg/projects/gstlal.html}},\
  \bibinfo {note} {accessed: 2015-07-01}\BibitemShut {NoStop}%
\bibitem [{\citenamefont {Dal~Canton}\ \emph {et~al.}(2014)\citenamefont
  {Dal~Canton}, \citenamefont {Nitz}, \citenamefont {Lundgren} \emph
  {et~al.}}]{pycbc1}%
  \BibitemOpen
  \bibfield  {author} {\bibinfo {author} {\bibfnamefont {T.}~\bibnamefont
  {Dal~Canton}}, \bibinfo {author} {\bibfnamefont {A.~H.}\ \bibnamefont
  {Nitz}}, \bibinfo {author} {\bibfnamefont {A.~P.}\ \bibnamefont {Lundgren}},
  \emph {et~al.},\ }\href {\doibase 10.1103/PhysRevD.90.082004} {\bibfield
  {journal} {\bibinfo  {journal} {Phys. Rev. D}\ }\textbf {\bibinfo {volume}
  {90}},\ \bibinfo {pages} {082004} (\bibinfo {year} {2014})}\BibitemShut
  {NoStop}%
\bibitem [{\citenamefont {Usman}\ \emph {et~al.}(2016)\citenamefont {Usman},
  \citenamefont {Nitz}, \citenamefont {Harry} \emph {et~al.}}]{pycbc2}%
  \BibitemOpen
  \bibfield  {author} {\bibinfo {author} {\bibfnamefont {S.}~\bibnamefont
  {Usman}}, \bibinfo {author} {\bibfnamefont {A.~H.}\ \bibnamefont {Nitz}},
  \bibinfo {author} {\bibfnamefont {I.~W.}\ \bibnamefont {Harry}},  \emph
  {et~al.},\ }\href {http://stacks.iop.org/0264-9381/33/i=21/a=215004}
  {\bibfield  {journal} {\bibinfo  {journal} {Classical and Quantum Gravity}\
  }\textbf {\bibinfo {volume} {33}},\ \bibinfo {pages} {215004} (\bibinfo
  {year} {2016})}\BibitemShut {NoStop}%
\bibitem [{\citenamefont {Nitz}\ \emph {et~al.}(2017)\citenamefont {Nitz} \emph
  {et~al.}}]{pycbc3}%
  \BibitemOpen
  \bibfield  {author} {\bibinfo {author} {\bibfnamefont {A.~H.}\ \bibnamefont
  {Nitz}} \emph {et~al.},\ }\href
  {https://zenodo.org/record/545845#.W59Xl5N_LBI} {\enquote {\bibinfo {title}
  {Pycbc software},}\ } (\bibinfo {year} {2017})\BibitemShut {NoStop}%
\bibitem [{\citenamefont {Nitz}(2018)}]{pycbc4}%
  \BibitemOpen
  \bibfield  {author} {\bibinfo {author} {\bibfnamefont {A.~H.}\ \bibnamefont
  {Nitz}},\ }\href {http://stacks.iop.org/0264-9381/35/i=3/a=035016} {\bibfield
   {journal} {\bibinfo  {journal} {Classical and Quantum Gravity}\ }\textbf
  {\bibinfo {volume} {35}},\ \bibinfo {pages} {035016} (\bibinfo {year}
  {2018})}\BibitemShut {NoStop}%
\bibitem [{\citenamefont {Adams}\ \emph {et~al.}(2016)\citenamefont {Adams},
  \citenamefont {Buskulic}, \citenamefont {Germain} \emph {et~al.}}]{MBTA}%
  \BibitemOpen
  \bibfield  {author} {\bibinfo {author} {\bibfnamefont {T.}~\bibnamefont
  {Adams}}, \bibinfo {author} {\bibfnamefont {D.}~\bibnamefont {Buskulic}},
  \bibinfo {author} {\bibfnamefont {V.}~\bibnamefont {Germain}},  \emph
  {et~al.},\ }\href {http://stacks.iop.org/0264-9381/33/i=17/a=175012}
  {\bibfield  {journal} {\bibinfo  {journal} {Classical and Quantum Gravity}\
  }\textbf {\bibinfo {volume} {33}},\ \bibinfo {pages} {175012} (\bibinfo
  {year} {2016})}\BibitemShut {NoStop}%
\bibitem [{\citenamefont {Abbott}\ \emph
  {et~al.}(2016{\natexlab{b}})\citenamefont {Abbott} \emph
  {et~al.}}]{CBCcompanion}%
  \BibitemOpen
  \bibfield  {author} {\bibinfo {author} {\bibfnamefont {B.~P.}\ \bibnamefont
  {Abbott}} \emph {et~al.} (\bibinfo {collaboration} {LIGO Scientific
  Collaboration and Virgo Collaboration}),\ }\href {\doibase
  10.1103/PhysRevD.93.122003} {\bibfield  {journal} {\bibinfo  {journal} {Phys.
  Rev. D}\ }\textbf {\bibinfo {volume} {93}},\ \bibinfo {pages} {122003}
  (\bibinfo {year} {2016}{\natexlab{b}})}\BibitemShut {NoStop}%
\bibitem [{\citenamefont {Abbott}\ \emph
  {et~al.}(2016{\natexlab{c}})\citenamefont {Abbott} \emph
  {et~al.}}]{O1BBHobs}%
  \BibitemOpen
  \bibfield  {author} {\bibinfo {author} {\bibfnamefont {B.~P.}\ \bibnamefont
  {Abbott}} \emph {et~al.} (\bibinfo {collaboration} {LIGO Scientific
  Collaboration and Virgo Collaboration}),\ }\href {\doibase
  10.1103/PhysRevX.6.041015} {\bibfield  {journal} {\bibinfo  {journal} {Phys.
  Rev. X}\ }\textbf {\bibinfo {volume} {6}},\ \bibinfo {pages} {041015}
  (\bibinfo {year} {2016}{\natexlab{c}})}\BibitemShut {NoStop}%
\bibitem [{\citenamefont {Dal~Canton}\ and\ \citenamefont
  {Harry}(2017)}]{O2Pycbc}%
  \BibitemOpen
  \bibfield  {author} {\bibinfo {author} {\bibfnamefont {T.}~\bibnamefont
  {Dal~Canton}}\ and\ \bibinfo {author} {\bibfnamefont {I.~W.}\ \bibnamefont
  {Harry}},\ }\href@noop {} {\  (\bibinfo {year} {2017})},\ \Eprint
  {http://arxiv.org/abs/1705.01845} {arXiv:1705.01845 [gr-qc]} \BibitemShut
  {NoStop}%
\bibitem [{\citenamefont {Abbott}\ \emph
  {et~al.}(2019{\natexlab{a}})\citenamefont {Abbott} \emph
  {et~al.}}]{O1-O2Catalog}%
  \BibitemOpen
  \bibfield  {author} {\bibinfo {author} {\bibfnamefont {B.~P.}\ \bibnamefont
  {Abbott}} \emph {et~al.} (\bibinfo {collaboration} {LIGO Scientific
  Collaboration and Virgo Collaboration}),\ }\href {\doibase
  10.1103/PhysRevX.9.031040} {\bibfield  {journal} {\bibinfo  {journal} {Phys. Rev. X}\ }\textbf {\bibinfo {volume} {9}},\ \bibinfo {pages} {031040}
  (\bibinfo {year} {2019}{\natexlab{b}})}\BibitemShut 
  {NoStop}%
\bibitem [{\citenamefont {Abbott}\ \emph
  {et~al.}(2017{\natexlab{e}})\citenamefont {Abbott} \emph {et~al.}}]{O1IMBH}%
  \BibitemOpen
  \bibfield  {author} {\bibinfo {author} {\bibfnamefont {B.~P.}\ \bibnamefont
  {Abbott}} \emph {et~al.} (\bibinfo {collaboration} {LIGO Scientific
  Collaboration and Virgo Collaboration}),\ }\href {\doibase
  10.1103/PhysRevD.96.022001} {\bibfield  {journal} {\bibinfo  {journal} {Phys.
  Rev. D}\ }\textbf {\bibinfo {volume} {96}},\ \bibinfo {pages} {022001}
  (\bibinfo {year} {2017}{\natexlab{e}})}\BibitemShut {NoStop}%
\bibitem [{\citenamefont {Rhoades}\ and\ \citenamefont
  {Ruffini}(1974)}]{Rhoades1974}%
  \BibitemOpen
  \bibfield  {author} {\bibinfo {author} {\bibfnamefont {C.~E.}\ \bibnamefont
  {Rhoades}}\ and\ \bibinfo {author} {\bibfnamefont {R.}~\bibnamefont
  {Ruffini}},\ }\href {\doibase 10.1103/PhysRevLett.32.324} {\bibfield
  {journal} {\bibinfo  {journal} {Phys. Rev. Lett.}\ }\textbf {\bibinfo
  {volume} {32}},\ \bibinfo {pages} {324} (\bibinfo {year} {1974})}\BibitemShut
  {NoStop}%
\bibitem [{\citenamefont {Kalogera}\ and\ \citenamefont
  {Baym}(1996)}]{Kalogera1996}%
  \BibitemOpen
  \bibfield  {author} {\bibinfo {author} {\bibfnamefont {V.}~\bibnamefont
  {Kalogera}}\ and\ \bibinfo {author} {\bibfnamefont {G.}~\bibnamefont
  {Baym}},\ }\href {http://stacks.iop.org/1538-4357/470/i=1/a=L61} {\bibfield
  {journal} {\bibinfo  {journal} {The Astrophysical Journal Letters}\ }\textbf
  {\bibinfo {volume} {470}},\ \bibinfo {pages} {L61} (\bibinfo {year}
  {1996})}\BibitemShut {NoStop}%
\bibitem [{\citenamefont {\"{O}zel}\ \emph {et~al.}(2012)\citenamefont
  {\"{O}zel}, \citenamefont {Psaltis}, \citenamefont {Narayan},\ and\
  \citenamefont {Villarreal}}]{OzelNSLower}%
  \BibitemOpen
  \bibfield  {author} {\bibinfo {author} {\bibfnamefont {F.}~\bibnamefont
  {\"{O}zel}}, \bibinfo {author} {\bibfnamefont {D.}~\bibnamefont {Psaltis}},
  \bibinfo {author} {\bibfnamefont {R.}~\bibnamefont {Narayan}}, \ and\
  \bibinfo {author} {\bibfnamefont {A.~S.}\ \bibnamefont {Villarreal}},\ }\href
  {http://stacks.iop.org/0004-637X/757/i=1/a=55} {\bibfield  {journal}
  {\bibinfo  {journal} {The Astrophysical Journal}\ }\textbf {\bibinfo {volume}
  {757}},\ \bibinfo {pages} {55} (\bibinfo {year} {2012})}\BibitemShut
  {NoStop}%
\bibitem [{\citenamefont {Lattimer}(2012)}]{Lattimer2012}%
  \BibitemOpen
  \bibfield  {author} {\bibinfo {author} {\bibfnamefont {J.~M.}\ \bibnamefont
  {Lattimer}},\ }\href {\doibase 10.1146/annurev-nucl-102711-095018} {\bibfield
   {journal} {\bibinfo  {journal} {Annual Review of Nuclear and Particle
  Science}\ }\textbf {\bibinfo {volume} {62}},\ \bibinfo {pages} {485}
  (\bibinfo {year} {2012})},\ \Eprint
  {http://arxiv.org/abs/https://doi.org/10.1146/annurev-nucl-102711-095018}
  {https://doi.org/10.1146/annurev-nucl-102711-095018} \BibitemShut {NoStop}%
\bibitem [{\citenamefont {Kiziltan}\ \emph {et~al.}(2013)\citenamefont
  {Kiziltan}, \citenamefont {Kottas}, \citenamefont {Yoreo},\ and\
  \citenamefont {Thorsett}}]{Kiziltan2013}%
  \BibitemOpen
  \bibfield  {author} {\bibinfo {author} {\bibfnamefont {B.}~\bibnamefont
  {Kiziltan}}, \bibinfo {author} {\bibfnamefont {A.}~\bibnamefont {Kottas}},
  \bibinfo {author} {\bibfnamefont {M.~D.}\ \bibnamefont {Yoreo}}, \ and\
  \bibinfo {author} {\bibfnamefont {S.~E.}\ \bibnamefont {Thorsett}},\ }\href
  {http://stacks.iop.org/0004-637X/778/i=1/a=66} {\bibfield  {journal}
  {\bibinfo  {journal} {The Astrophysical Journal}\ }\textbf {\bibinfo {volume}
  {778}},\ \bibinfo {pages} {66} (\bibinfo {year} {2013})}\BibitemShut
  {NoStop}%
\bibitem [{\citenamefont {Linares}\ \emph {et~al.}(2018)\citenamefont
  {Linares}, \citenamefont {Shahbaz},\ and\ \citenamefont
  {Casares}}]{MassiveNS}%
  \BibitemOpen
  \bibfield  {author} {\bibinfo {author} {\bibfnamefont {M.}~\bibnamefont
  {Linares}}, \bibinfo {author} {\bibfnamefont {T.}~\bibnamefont {Shahbaz}}, \
  and\ \bibinfo {author} {\bibfnamefont {J.}~\bibnamefont {Casares}},\ }\href
  {http://stacks.iop.org/0004-637X/859/i=1/a=54} {\bibfield  {journal}
  {\bibinfo  {journal} {The Astrophysical Journal}\ }\textbf {\bibinfo {volume}
  {859}},\ \bibinfo {pages} {54} (\bibinfo {year} {2018})}\BibitemShut
  {NoStop}%
\bibitem [{\citenamefont {\"{O}zel}\ and\ \citenamefont
  {Freire}(2016)}]{Ozel2016}%
  \BibitemOpen
  \bibfield  {author} {\bibinfo {author} {\bibfnamefont {F.}~\bibnamefont
  {\"{O}zel}}\ and\ \bibinfo {author} {\bibfnamefont {P.}~\bibnamefont
  {Freire}},\ }\href {\doibase 10.1146/annurev-astro-081915-023322} {\bibfield
  {journal} {\bibinfo  {journal} {Annual Review of Astronomy and Astrophysics}\
  }\textbf {\bibinfo {volume} {54}},\ \bibinfo {pages} {401} (\bibinfo {year}
  {2016})},\ \Eprint
  {http://arxiv.org/abs/https://doi.org/10.1146/annurev-astro-081915-023322}
  {https://doi.org/10.1146/annurev-astro-081915-023322} \BibitemShut {NoStop}%
 \bibitem [{\citenamefont {P\"{u}rrer} (2014)}]{SEOBNR_Double}%
  \BibitemOpen
  \bibfield  {author} {\bibinfo {author} {\bibfnamefont {M}~\bibnamefont
  {P\"{u}rrer}},\ }\href {\doibase 10.1088/0264-9381/31/19/195010} {\bibfield
  {journal} {\bibinfo  {journal} {Classical and Quantum Gravity}\
  }\textbf {\bibinfo {volume} {31}},\ \bibinfo {pages} {195010} (\bibinfo {year}
  {2014})}\BibitemShut {NoStop}%
\bibitem [{\citenamefont {O'Shaughnessy}\ \emph {et~al.}(2005)\citenamefont
  {O'Shaughnessy}, \citenamefont {Kaplan}, \citenamefont {Kalogera},\ and\
  \citenamefont {Belczynski}}]{Shaughnessy2005}%
  \BibitemOpen
  \bibfield  {author} {\bibinfo {author} {\bibfnamefont {R.}~\bibnamefont
  {O'Shaughnessy}}, \bibinfo {author} {\bibfnamefont {J.}~\bibnamefont
  {Kaplan}}, \bibinfo {author} {\bibfnamefont {V.}~\bibnamefont {Kalogera}}, \
  and\ \bibinfo {author} {\bibfnamefont {K.}~\bibnamefont {Belczynski}},\
  }\href {http://stacks.iop.org/0004-637X/632/i=2/a=1035} {\bibfield  {journal}
  {\bibinfo  {journal} {The Astrophysical Journal}\ }\textbf {\bibinfo {volume}
  {632}},\ \bibinfo {pages} {1035} (\bibinfo {year} {2005})}\BibitemShut
  {NoStop}%
\bibitem [{\citenamefont {Belczynski}\ \emph {et~al.}(2014)\citenamefont
  {Belczynski}, \citenamefont {Buonanno}, \citenamefont {Cantiello},
  \citenamefont {Fryer}, \citenamefont {Holz}, \citenamefont {Mandel},
  \citenamefont {Miller},\ and\ \citenamefont {Walczak}}]{Belczynski2014}%
  \BibitemOpen
  \bibfield  {author} {\bibinfo {author} {\bibfnamefont {K.}~\bibnamefont
  {Belczynski}}, \bibinfo {author} {\bibfnamefont {A.}~\bibnamefont
  {Buonanno}}, \bibinfo {author} {\bibfnamefont {M.}~\bibnamefont {Cantiello}},
  \bibinfo {author} {\bibfnamefont {C.~L.}\ \bibnamefont {Fryer}}, \bibinfo
  {author} {\bibfnamefont {D.~E.}\ \bibnamefont {Holz}}, \bibinfo {author}
  {\bibfnamefont {I.}~\bibnamefont {Mandel}}, \bibinfo {author} {\bibfnamefont
  {M.~C.}\ \bibnamefont {Miller}}, \ and\ \bibinfo {author} {\bibfnamefont
  {M.}~\bibnamefont {Walczak}},\ }\href
  {http://stacks.iop.org/0004-637X/789/i=2/a=120} {\bibfield  {journal}
  {\bibinfo  {journal} {The Astrophysical Journal}\ }\textbf {\bibinfo {volume}
  {789}},\ \bibinfo {pages} {120} (\bibinfo {year} {2014})}\BibitemShut
  {NoStop}%
\bibitem [{\citenamefont {de~Mink}\ and\ \citenamefont
  {Belczynski}(2015)}]{deMink2015}%
  \BibitemOpen
  \bibfield  {author} {\bibinfo {author} {\bibfnamefont {S.~E.}\ \bibnamefont
  {de~Mink}}\ and\ \bibinfo {author} {\bibfnamefont {K.}~\bibnamefont
  {Belczynski}},\ }\href {http://stacks.iop.org/0004-637X/814/i=1/a=58}
  {\bibfield  {journal} {\bibinfo  {journal} {The Astrophysical Journal}\
  }\textbf {\bibinfo {volume} {814}},\ \bibinfo {pages} {58} (\bibinfo {year}
  {2015})}\BibitemShut {NoStop}%
\bibitem [{\citenamefont {Miller}\ and\ \citenamefont
  {Colbert}(2004)}]{Miller2004}%
  \BibitemOpen
  \bibfield  {author} {\bibinfo {author} {\bibfnamefont {M.~C.}\ \bibnamefont
  {Miller}}\ and\ \bibinfo {author} {\bibfnamefont {E.~J.~M.}\ \bibnamefont
  {Colbert}},\ }\href {\doibase 10.1142/S0218271804004426} {\bibfield
  {journal} {\bibinfo  {journal} {Int. J. Mod. Phys.}\ }\textbf {\bibinfo
  {volume} {D13}},\ \bibinfo {pages} {1} (\bibinfo {year} {2004})},\ \Eprint
  {http://arxiv.org/abs/astro-ph/0308402} {arXiv:astro-ph/0308402 [astro-ph]}
  \BibitemShut {NoStop}%
\bibitem [{\citenamefont {Abbott}\ \emph {et~al.}(2018)\citenamefont {Abbott}
  \emph {et~al.}}]{Subsolar2018}%
  \BibitemOpen
  \bibfield  {author} {\bibinfo {author} {\bibfnamefont {B.~P.}\ \bibnamefont
  {Abbott}} \emph {et~al.} (\bibinfo {collaboration} {Virgo, LIGO
  Scientific}),\ }\href {\doibase
  10.1103/PhysRevLett.121.231103} {\bibfield  {journal} {\bibinfo  {journal} {Phys. Rev. Lett.}\ }\textbf {\bibinfo {volume} {121}},\ \bibinfo {pages} {231103}
  (\bibinfo {year} {2018}{\natexlab{b}})}\BibitemShut 
  {NoStop}%
\bibitem [{\citenamefont {{Hessels}}\ \emph {et~al.}(2006)\citenamefont
  {{Hessels}}, \citenamefont {{Ransom}}, \citenamefont {{Stairs}},
  \citenamefont {{Freire}}, \citenamefont {{Kaspi}},\ and\ \citenamefont
  {{Camilo}}}]{Hessels2006}%
  \BibitemOpen
  \bibfield  {author} {\bibinfo {author} {\bibfnamefont {J.~W.~T.}\
  \bibnamefont {{Hessels}}}, \bibinfo {author} {\bibfnamefont {S.~M.}\
  \bibnamefont {{Ransom}}}, \bibinfo {author} {\bibfnamefont {I.~H.}\
  \bibnamefont {{Stairs}}}, \bibinfo {author} {\bibfnamefont {P.~C.~C.}\
  \bibnamefont {{Freire}}}, \bibinfo {author} {\bibfnamefont {V.~M.}\
  \bibnamefont {{Kaspi}}}, \ and\ \bibinfo {author} {\bibfnamefont
  {F.}~\bibnamefont {{Camilo}}},\ }\href {\doibase 10.1126/science.1123430}
  {\bibfield  {journal} {\bibinfo  {journal} {Science}\ }\textbf {\bibinfo
  {volume} {311}},\ \bibinfo {pages} {1901} (\bibinfo {year} {2006})},\ \Eprint
  {http://arxiv.org/abs/astro-ph/0601337} {astro-ph/0601337} \BibitemShut
  {NoStop}%
\bibitem [{\citenamefont {Kramer}\ and\ \citenamefont
  {Wex}(2009)}]{Kramer2009}%
  \BibitemOpen
  \bibfield  {author} {\bibinfo {author} {\bibfnamefont {M.}~\bibnamefont
  {Kramer}}\ and\ \bibinfo {author} {\bibfnamefont {N.}~\bibnamefont {Wex}},\
  }\href {http://stacks.iop.org/0264-9381/26/i=7/a=073001} {\bibfield
  {journal} {\bibinfo  {journal} {Classical and Quantum Gravity}\ }\textbf
  {\bibinfo {volume} {26}},\ \bibinfo {pages} {073001} (\bibinfo {year}
  {2009})}\BibitemShut {NoStop}%
\bibitem [{\citenamefont {{Fabian}}\ \emph {et~al.}(2012)\citenamefont
  {{Fabian}}, \citenamefont {{Wilkins}}, \citenamefont {{Miller}},
  \citenamefont {{Reis}}, \citenamefont {{Reynolds}}, \citenamefont
  {{Cackett}}, \citenamefont {{Nowak}}, \citenamefont {{Pooley}}, \citenamefont
  {{Pottschmidt}}, \citenamefont {{Sanders}}, \citenamefont {{Ross}},\ and\
  \citenamefont {{Wilms}}}]{Fabian2012}%
  \BibitemOpen
  \bibfield  {author} {\bibinfo {author} {\bibfnamefont {A.~C.}\ \bibnamefont
  {{Fabian}}}, \bibinfo {author} {\bibfnamefont {D.~R.}\ \bibnamefont
  {{Wilkins}}}, \bibinfo {author} {\bibfnamefont {J.~M.}\ \bibnamefont
  {{Miller}}}, \bibinfo {author} {\bibfnamefont {R.~C.}\ \bibnamefont
  {{Reis}}}, \bibinfo {author} {\bibfnamefont {C.~S.}\ \bibnamefont
  {{Reynolds}}}, \bibinfo {author} {\bibfnamefont {E.~M.}\ \bibnamefont
  {{Cackett}}}, \bibinfo {author} {\bibfnamefont {M.~A.}\ \bibnamefont
  {{Nowak}}}, \bibinfo {author} {\bibfnamefont {G.~G.}\ \bibnamefont
  {{Pooley}}}, \bibinfo {author} {\bibfnamefont {K.}~\bibnamefont
  {{Pottschmidt}}}, \bibinfo {author} {\bibfnamefont {J.~S.}\ \bibnamefont
  {{Sanders}}}, \bibinfo {author} {\bibfnamefont {R.~R.}\ \bibnamefont
  {{Ross}}}, \ and\ \bibinfo {author} {\bibfnamefont {J.}~\bibnamefont
  {{Wilms}}},\ }\href {\doibase 10.1111/j.1365-2966.2012.21185.x} {\bibfield
  {journal} {\bibinfo  {journal} {Monthly Notices of the Royal Astronomical
  Society: Letters}\ }\textbf {\bibinfo {volume} {424}},\ \bibinfo {pages}
  {217} (\bibinfo {year} {2012})},\ \Eprint {http://arxiv.org/abs/1204.5854}
  {arXiv:1204.5854 [astro-ph.HE]} \BibitemShut {NoStop}%
\bibitem [{\citenamefont {{Gou}}\ \emph {et~al.}(2011)\citenamefont {{Gou}},
  \citenamefont {{McClintock}}, \citenamefont {{Reid}}, \citenamefont
  {{Orosz}}, \citenamefont {{Steiner}}, \citenamefont {{Narayan}},
  \citenamefont {{Xiang}}, \citenamefont {{Remillard}}, \citenamefont
  {{Arnaud}},\ and\ \citenamefont {{Davis}}}]{Gou2011}%
  \BibitemOpen
  \bibfield  {author} {\bibinfo {author} {\bibfnamefont {L.}~\bibnamefont
  {{Gou}}}, \bibinfo {author} {\bibfnamefont {J.~E.}\ \bibnamefont
  {{McClintock}}}, \bibinfo {author} {\bibfnamefont {M.~J.}\ \bibnamefont
  {{Reid}}}, \bibinfo {author} {\bibfnamefont {J.~A.}\ \bibnamefont {{Orosz}}},
  \bibinfo {author} {\bibfnamefont {J.~F.}\ \bibnamefont {{Steiner}}}, \bibinfo
  {author} {\bibfnamefont {R.}~\bibnamefont {{Narayan}}}, \bibinfo {author}
  {\bibfnamefont {J.}~\bibnamefont {{Xiang}}}, \bibinfo {author} {\bibfnamefont
  {R.~A.}\ \bibnamefont {{Remillard}}}, \bibinfo {author} {\bibfnamefont
  {K.~A.}\ \bibnamefont {{Arnaud}}}, \ and\ \bibinfo {author} {\bibfnamefont
  {S.~W.}\ \bibnamefont {{Davis}}},\ }\href {\doibase
  10.1088/0004-637X/742/2/85} {\bibfield  {journal} {\bibinfo  {journal}
  {\apj}\ }\textbf {\bibinfo {volume} {742}},\ \bibinfo {eid} {85} (\bibinfo
  {year} {2011})},\ \Eprint {http://arxiv.org/abs/1106.3690} {arXiv:1106.3690
  [astro-ph.HE]} \BibitemShut {NoStop}%
\bibitem [{\citenamefont {McClintock}\ \emph {et~al.}(2011)\citenamefont
  {McClintock}, \citenamefont {Narayan}, \citenamefont {Davis}, \citenamefont
  {Gou}, \citenamefont {Kulkarni}, \citenamefont {Orosz}, \citenamefont
  {Penna}, \citenamefont {Remillard},\ and\ \citenamefont
  {Steiner}}]{Mcclintock2011}%
  \BibitemOpen
  \bibfield  {author} {\bibinfo {author} {\bibfnamefont {J.~E.}\ \bibnamefont
  {McClintock}}, \bibinfo {author} {\bibfnamefont {R.}~\bibnamefont {Narayan}},
  \bibinfo {author} {\bibfnamefont {S.~W.}\ \bibnamefont {Davis}}, \bibinfo
  {author} {\bibfnamefont {L.}~\bibnamefont {Gou}}, \bibinfo {author}
  {\bibfnamefont {A.}~\bibnamefont {Kulkarni}}, \bibinfo {author}
  {\bibfnamefont {J.~A.}\ \bibnamefont {Orosz}}, \bibinfo {author}
  {\bibfnamefont {R.~F.}\ \bibnamefont {Penna}}, \bibinfo {author}
  {\bibfnamefont {R.~A.}\ \bibnamefont {Remillard}}, \ and\ \bibinfo {author}
  {\bibfnamefont {J.~F.}\ \bibnamefont {Steiner}},\ }\href
  {http://stacks.iop.org/0264-9381/28/i=11/a=114009} {\bibfield  {journal}
  {\bibinfo  {journal} {Classical and Quantum Gravity}\ }\textbf {\bibinfo
  {volume} {28}},\ \bibinfo {pages} {114009} (\bibinfo {year}
  {2011})}\BibitemShut {NoStop}%
\bibitem [{\citenamefont {Misner}\ \emph {et~al.}(1973)\citenamefont {Misner},
  \citenamefont {Thorne},\ and\ \citenamefont {Wheeler}}]{MisnerGrav}%
  \BibitemOpen
  \bibfield  {author} {\bibinfo {author} {\bibfnamefont {C.~W.}\ \bibnamefont
  {Misner}}, \bibinfo {author} {\bibfnamefont {K.~S.}\ \bibnamefont {Thorne}},
  \ and\ \bibinfo {author} {\bibfnamefont {J.~A.}\ \bibnamefont {Wheeler}},\
  }\href@noop {} {\emph {\bibinfo {title} {Gravitation}}}\ (\bibinfo
  {publisher} {W. H. Freeman and Company},\ \bibinfo {year} {1973})\BibitemShut
  {NoStop}%
\bibitem [{\citenamefont {Privitera}\ \emph {et~al.}(2014)\citenamefont
  {Privitera}, \citenamefont {Mohapatra}, \citenamefont {Ajith}, \citenamefont
  {Cannon}, \citenamefont {Fotopoulos}, \citenamefont {Frei}, \citenamefont
  {Hanna}, \citenamefont {Weinstein},\ and\ \citenamefont
  {Whelan}}]{privitera2014improving}%
  \BibitemOpen
  \bibfield  {author} {\bibinfo {author} {\bibfnamefont {S.}~\bibnamefont
  {Privitera}}, \bibinfo {author} {\bibfnamefont {S.~R.~P.}\ \bibnamefont
  {Mohapatra}}, \bibinfo {author} {\bibfnamefont {P.}~\bibnamefont {Ajith}},
  \bibinfo {author} {\bibfnamefont {K.}~\bibnamefont {Cannon}}, \bibinfo
  {author} {\bibfnamefont {N.}~\bibnamefont {Fotopoulos}}, \bibinfo {author}
  {\bibfnamefont {M.~A.}\ \bibnamefont {Frei}}, \bibinfo {author}
  {\bibfnamefont {C.}~\bibnamefont {Hanna}}, \bibinfo {author} {\bibfnamefont
  {A.~J.}\ \bibnamefont {Weinstein}}, \ and\ \bibinfo {author} {\bibfnamefont
  {J.~T.}\ \bibnamefont {Whelan}},\ }\href@noop {} {\bibfield  {journal}
  {\bibinfo  {journal} {Physical Review D}\ }\textbf {\bibinfo {volume} {89}},\
  \bibinfo {pages} {024003} (\bibinfo {year} {2014})}\BibitemShut {NoStop}%
\bibitem [{\citenamefont {Harry}\ \emph {et~al.}(2014)\citenamefont {Harry},
  \citenamefont {Nitz}, \citenamefont {Brown}, \citenamefont {Lundgren},
  \citenamefont {Ochsner},\ and\ \citenamefont {Keppel}}]{fittingfactor}%
  \BibitemOpen
  \bibfield  {author} {\bibinfo {author} {\bibfnamefont {I.~W.}\ \bibnamefont
  {Harry}}, \bibinfo {author} {\bibfnamefont {A.~H.}\ \bibnamefont {Nitz}},
  \bibinfo {author} {\bibfnamefont {D.~A.}\ \bibnamefont {Brown}}, \bibinfo
  {author} {\bibfnamefont {A.~P.}\ \bibnamefont {Lundgren}}, \bibinfo {author}
  {\bibfnamefont {E.}~\bibnamefont {Ochsner}}, \ and\ \bibinfo {author}
  {\bibfnamefont {D.}~\bibnamefont {Keppel}},\ }\href {\doibase
  10.1103/PhysRevD.89.024010} {\bibfield  {journal} {\bibinfo  {journal} {Phys.
  Rev. D}\ }\textbf {\bibinfo {volume} {89}},\ \bibinfo {pages} {024010}
  (\bibinfo {year} {2014})}\BibitemShut {NoStop}%
\bibitem [{\citenamefont {Apostolatos}(1995)}]{FFdef}%
  \BibitemOpen
  \bibfield  {author} {\bibinfo {author} {\bibfnamefont {T.~A.}\ \bibnamefont
  {Apostolatos}},\ }\href {\doibase 10.1103/PhysRevD.52.605} {\bibfield
  {journal} {\bibinfo  {journal} {Phys. Rev. D}\ }\textbf {\bibinfo {volume}
  {52}},\ \bibinfo {pages} {605} (\bibinfo {year} {1995})}\BibitemShut
  {NoStop}%
\bibitem [{psd(2017)}]{psd}%
  \BibitemOpen
  \href@noop {} {\enquote {\bibinfo {title} {aligo sensitivity projections for
  o2},}\ }\bibinfo {howpublished} {\url{https://dcc.ligo.org/LIGO-T1600302}}
  (\bibinfo {year} {2017}),\ \bibinfo {note} {dcc link}\BibitemShut {NoStop}%
\bibitem [{\citenamefont {Ajith}\ \emph {et~al.}(2014)\citenamefont {Ajith},
  \citenamefont {Fotopoulos}, \citenamefont {Privitera}, \citenamefont
  {Neunzert}, \citenamefont {Mazumder},\ and\ \citenamefont
  {Weinstein}}]{ajith2014effectual}%
  \BibitemOpen
  \bibfield  {author} {\bibinfo {author} {\bibfnamefont {P.}~\bibnamefont
  {Ajith}}, \bibinfo {author} {\bibfnamefont {N.}~\bibnamefont {Fotopoulos}},
  \bibinfo {author} {\bibfnamefont {S.}~\bibnamefont {Privitera}}, \bibinfo
  {author} {\bibfnamefont {A.}~\bibnamefont {Neunzert}}, \bibinfo {author}
  {\bibfnamefont {N.}~\bibnamefont {Mazumder}}, \ and\ \bibinfo {author}
  {\bibfnamefont {A.~J.}\ \bibnamefont {Weinstein}},\ }\href@noop {} {\bibfield
   {journal} {\bibinfo  {journal} {Physical Review D}\ }\textbf {\bibinfo
  {volume} {89}},\ \bibinfo {pages} {084041} (\bibinfo {year}
  {2014})}\BibitemShut {NoStop}%
\bibitem [{\citenamefont {Boh\'e}\ \emph {et~al.}(2017)\citenamefont {Boh\'e},
  \citenamefont {Shao}, \citenamefont {Taracchini} \emph
  {et~al.}}]{SEOBNRv4ROM}%
  \BibitemOpen
  \bibfield  {author} {\bibinfo {author} {\bibfnamefont {A.}~\bibnamefont
  {Boh\'e}}, \bibinfo {author} {\bibfnamefont {L.}~\bibnamefont {Shao}},
  \bibinfo {author} {\bibfnamefont {A.}~\bibnamefont {Taracchini}},  \emph
  {et~al.},\ }\href {\doibase 10.1103/PhysRevD.95.044028} {\bibfield  {journal}
  {\bibinfo  {journal} {Phys. Rev. D}\ }\textbf {\bibinfo {volume} {95}},\
  \bibinfo {pages} {044028} (\bibinfo {year} {2017})}\BibitemShut {NoStop}%
  \bibitem [{\citenamefont {Hannam}\ \emph {et~al.}(2014)\citenamefont {Hannam},
  \citenamefont {Schmidt}, \citenamefont {Boh\'e} \emph
  {et~al.}}]{IMRPhenomPv2}%
  \BibitemOpen
  \bibfield  {author} {\bibinfo {author} {\bibfnamefont {M.}~\bibnamefont
  {Hannam}}, \bibinfo {author} {\bibfnamefont {P.}~\bibnamefont {Schmidt}},
  \bibinfo {author} {\bibfnamefont {A.}~\bibnamefont {Boh\'e}},  \emph
  {et~al.},\ }\href {\doibase 10.1103/PhysRevLett.113.151101} {\bibfield  {journal}
  {\bibinfo  {journal} {Phys. Rev. Lett.}\ }\textbf {\bibinfo {volume} {113}},\
  \bibinfo {pages} {151101} (\bibinfo {year} {2014})}\BibitemShut {NoStop}%
\bibitem [{\citenamefont {Buonanno}\ \emph {et~al.}(2009)\citenamefont
  {Buonanno}, \citenamefont {Iyer}, \citenamefont {Ochsner}, \citenamefont
  {Pan},\ and\ \citenamefont {Sathyaprakash}}]{TaylorF2}%
  \BibitemOpen
  \bibfield  {author} {\bibinfo {author} {\bibfnamefont {A.}~\bibnamefont
  {Buonanno}}, \bibinfo {author} {\bibfnamefont {B.~R.}\ \bibnamefont {Iyer}},
  \bibinfo {author} {\bibfnamefont {E.}~\bibnamefont {Ochsner}}, \bibinfo
  {author} {\bibfnamefont {Y.}~\bibnamefont {Pan}}, \ and\ \bibinfo {author}
  {\bibfnamefont {B.~S.}\ \bibnamefont {Sathyaprakash}},\ }\href {\doibase
  10.1103/PhysRevD.80.084043} {\bibfield  {journal} {\bibinfo  {journal} {Phys.
  Rev. D}\ }\textbf {\bibinfo {volume} {80}},\ \bibinfo {pages} {084043}
  (\bibinfo {year} {2009})}\BibitemShut {NoStop}%
\bibitem [{\citenamefont {Brown}\ \emph {et~al.}(2012)\citenamefont {Brown},
  \citenamefont {Harry}, \citenamefont {Lundgren},\ and\ \citenamefont
  {Nitz}}]{geometric}%
  \BibitemOpen
  \bibfield  {author} {\bibinfo {author} {\bibfnamefont {D.~A.}\ \bibnamefont
  {Brown}}, \bibinfo {author} {\bibfnamefont {I.}~\bibnamefont {Harry}},
  \bibinfo {author} {\bibfnamefont {A.}~\bibnamefont {Lundgren}}, \ and\
  \bibinfo {author} {\bibfnamefont {A.~H.}\ \bibnamefont {Nitz}},\ }\href
  {\doibase 10.1103/PhysRevD.86.084017} {\bibfield  {journal} {\bibinfo
  {journal} {Phys. Rev. D}\ }\textbf {\bibinfo {volume} {86}},\ \bibinfo
  {pages} {084017} (\bibinfo {year} {2012})}\BibitemShut {NoStop}%
\bibitem [{\citenamefont {Harry}\ \emph {et~al.}(2009)\citenamefont {Harry},
  \citenamefont {Allen},\ and\ \citenamefont {Sathyaprakash}}]{stochastic}%
  \BibitemOpen
  \bibfield  {author} {\bibinfo {author} {\bibfnamefont {I.~W.}\ \bibnamefont
  {Harry}}, \bibinfo {author} {\bibfnamefont {B.}~\bibnamefont {Allen}}, \ and\
  \bibinfo {author} {\bibfnamefont {B.~S.}\ \bibnamefont {Sathyaprakash}},\
  }\href {\doibase 10.1103/PhysRevD.80.104014} {\bibfield  {journal} {\bibinfo
  {journal} {Phys. Rev. D}\ }\textbf {\bibinfo {volume} {80}},\ \bibinfo
  {pages} {104014} (\bibinfo {year} {2009})}\BibitemShut {NoStop}%
\bibitem [{O2M(2018)}]{O2Methods}%
  \BibitemOpen
  \href@noop {} {\enquote {\bibinfo {title} {The gstlal search analysis methods
  for compact binary mergers in advanced ligo's second and advanced virgo's
  first observing runs},}\ }\bibinfo {howpublished}
  {\url{https://dcc.ligo.org/LIGO-P1700411}} (\bibinfo {year} {2018}),\
  \bibinfo {note} {dcc link}\BibitemShut {NoStop}%
\bibitem [{\citenamefont {Marsat}\ \emph {et~al.}(2014)\citenamefont {Marsat},
  \citenamefont {Boh{\'e}}, \citenamefont {Blanchet},\ and\ \citenamefont
  {Buonanno}}]{SpinTaylorT4}%
  \BibitemOpen
  \bibfield  {author} {\bibinfo {author} {\bibfnamefont {S.}~\bibnamefont
  {Marsat}}, \bibinfo {author} {\bibfnamefont {A.}~\bibnamefont {Boh{\'e}}},
  \bibinfo {author} {\bibfnamefont {L.}~\bibnamefont {Blanchet}}, \ and\
  \bibinfo {author} {\bibfnamefont {A.}~\bibnamefont {Buonanno}},\ }\href
  {http://stacks.iop.org/0264-9381/31/i=2/a=025023} {\bibfield  {journal}
  {\bibinfo  {journal} {Classical and Quantum Gravity}\ }\textbf {\bibinfo
  {volume} {31}},\ \bibinfo {pages} {025023} (\bibinfo {year}
  {2014})}\BibitemShut {NoStop}%
\bibitem [{\citenamefont {Pan}\ \emph {et~al.}(2011)\citenamefont {Pan},
  \citenamefont {Buonanno}, \citenamefont {Boyle}, \citenamefont {Buchman}
  \emph {et~al.}}]{EOBNRv2HM}%
  \BibitemOpen
  \bibfield  {author} {\bibinfo {author} {\bibfnamefont {Y.}~\bibnamefont
  {Pan}}, \bibinfo {author} {\bibfnamefont {A.}~\bibnamefont {Buonanno}},
  \bibinfo {author} {\bibfnamefont {M.}~\bibnamefont {Boyle}}, \bibinfo
  {author} {\bibfnamefont {L.~T.}\ \bibnamefont {Buchman}},  \emph {et~al.},\
  }\href {\doibase 10.1103/PhysRevD.84.124052} {\bibfield  {journal} {\bibinfo
  {journal} {Phys. Rev. D}\ }\textbf {\bibinfo {volume} {84}},\ \bibinfo
  {pages} {124052} (\bibinfo {year} {2011})}\BibitemShut {NoStop}%
\bibitem [{\citenamefont {Calder\'on~Bustillo}\ \emph
  {et~al.}(2018)\citenamefont {Calder\'on~Bustillo}, \citenamefont {Salemi},
  \citenamefont {Dal~Canton},\ and\ \citenamefont {Jani}}]{IMBHHM}%
  \BibitemOpen
  \bibfield  {author} {\bibinfo {author} {\bibfnamefont {J.}~\bibnamefont
  {Calder\'on~Bustillo}}, \bibinfo {author} {\bibfnamefont {F.}~\bibnamefont
  {Salemi}}, \bibinfo {author} {\bibfnamefont {T.}~\bibnamefont {Dal~Canton}},
  \ and\ \bibinfo {author} {\bibfnamefont {K.~P.}\ \bibnamefont {Jani}},\
  }\href {\doibase 10.1103/PhysRevD.97.024016} {\bibfield  {journal} {\bibinfo
  {journal} {Phys. Rev. D}\ }\textbf {\bibinfo {volume} {97}},\ \bibinfo
  {pages} {024016} (\bibinfo {year} {2018})}\BibitemShut {NoStop}%
\end{thebibliography}

%

\end{document}